# Challenges in Plasmonic Catalysis

Emiliano Cortés[1*], Lucas V. Besteiro[2], Alessandro Alabastri[3], Andrea Baldi[4,5],

Giulia Tagliabue[6], Angela Demetriadou[7], and Prineha Narang[8]

1- Chair in Hybrid Nanosystems, Nanoinstitute Munich, Faculty of Physics, Ludwig-Maximilians-Universität München, 80539 München, Germany

2- CINBIO, Universidade de Vigo, 36310 Vigo, Spain

3- Department of Electrical and Computer Engineering, Rice University, 6100 Main Street MS-378, Houston, Texas 77005, United States

4- DIFFER - Dutch Institute for Fundamental Energy Research, De Zaale 20, 5612 AJ Eindhoven, The Netherlands

5- Department of Physics and Astronomy, Vrije Universiteit Amsterdam, De Boelelaan 1081, 1081 HV Amsterdam, The Netherlands

6- Laboratory of Nanoscience for Energy Technologies (LNET), EPFL, 1015 Lausanne, Switzerland

7- School of Physics and Astronomy, University of Birmingham, Birmingham B15 2TT, United Kingdom

8- John A. Paulson School of Engineering and Applied Sciences, Harvard University, Cambridge, Massachusetts 02138, United States

Corresponding author E.C: Emiliano.Cortes@lmu.de

**Abstract**

The use of nanoplasmonics to control light and heat close to the thermodynamic limit enables exciting opportunities in the field of plasmonic catalysis. The decay of plasmonic excitations creates highly nonequilibrium distributions of hot carriers that can initiate or catalyze reactions through both thermal and nonthermal pathways. In this Perspective, we present the current understanding in the field of plasmonic catalysis, capturing vibrant debates in the literature, and discuss future avenues of exploration to overcome critical bottlenecks. Our Perspective spans first-principles theory and computation of correlated and far-from-equilibrium light–matter interactions, synthesis of new nanoplasmonic hybrids, and new steady-state and ultrafast spectroscopic probes of interactions in plasmonic catalysis, recognizing the key contributions of each discipline in realizing the promise of plasmonic catalysis. We conclude with our vision for fundamental and technological advances in the field of plasmon-driven chemical reactions in the coming years.

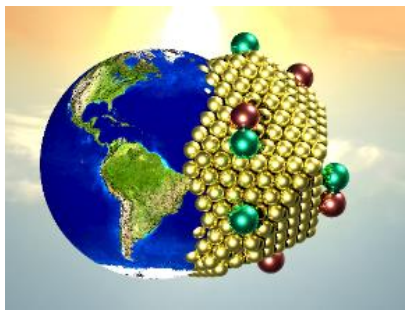

**Keywords:** nanoplasmonics, nanochemistry, photocatalysis, nonequilibrium dynamics, hot carriers

Chemical reactions are ubiquitous and play critical roles in everyday life. For decades, chemists have increased the rate of chemical reactions using metal surfaces, enzymes, molecular catalysts, or photocatalysts. However, traditional catalysts have well known limitations regarding reactivity, selectivity, and/or stability. Moreover, for most current industrial catalytic processes, the catalysts require high temperatures and/or pressures to operate efficiently. An emerging field in catalysis is the use of plasmon resonances in metal nanoparticles (NPs) to control the rate and selectivity of photocatalytic reactions. This research sits at the interface between chemistry, plasmonics, and cavity quantum electrodynamics, and has opened a new avenue for catalysis, leveraging the excitation of carriers in irradiated metal NPs as well as the quantum nature of light–molecule interactions to drive chemical reactions. The control of light and heat close to the thermodynamic limit enables exciting opportunities to explore the nascent field of plasmonic catalysis. Whereas most reactions obey Kasha's rule, in which photochemistry proceeds from the lowest excited energy states regardless of excitation wavelength, concepts from the fields of plasmonic chemistry and quantum optics present a pathway to correlated light–matter interactions that overcome conventional limits in chemical dynamics and quantum chemistry. Indeed, the fact that photoexcited carriers can affect the reactivity of molecules adsorbed on metal surfaces has been known for quite some time.[1,2] However, these transformations typically required high-power pulsed laser excitations, which is far from the dream of sunlight-driven chemical transformations.[3-6]

In recent years, the idea of using the energy gathered by plasmon excitations in nanomaterials for catalyzing chemical reactions has garnered considerable interest.[7] Here, we discuss the current understanding and future outlook of the field from an interdisciplinary perspective, spanning across first-principles theory and the computation of correlated and nonequilibrium light–matter interactions, synthesis of nanoplasmonic hybrids, and spectroscopies and probes of interactions in plasmonic catalysis. In this Perspective, we highlight the remarkable advances the field has made juxtaposed with the plethora of open questions.

Our Perspective is organized as follows. We begin with a discussion of predictions and theoretical approaches to describe plasmonic systems out of equilibrium. Next, we present a discussion of thermal effects in plasmonic catalysis, underlining the advances in theoretical and computational descriptions of these fields. Building on these concepts, we present spectroscopic and chemical signatures of plasmon-induced transformations, highlighting the convolution of direct charge transfer and strongly temperature-dependent effects. The collection of plasmonically excited carriers forms the subject of the next two sections, with an in-depth presentation of different interfacial approaches, including plasmonic metal-semiconductor junctions and metal–metal and metal–molecule hybrids. We further discuss plasmonic chemistry at the picoscale, highlighting the atomic nature of chemical reactivity. We end the discussion section by presenting the incipient and promising intersection of plasmonics with strong-coupling chemistry. Finally, we present our vision of the current bottlenecks in the field of plasmon-driven catalysis and propose future directions to pursue.

**Predicting and Understanding Nonequilibrium Nanoplasmonics at the Microscopic Level**

Predictions of the optical response of complex photonic structures have relied for decades on (semi)empirical constitutive parameters, namely the material permittivity and permeability. However, as nanophotonics approaches the atomic limit, the predominantly empirical nature of these dielectric functions renders a number of questions unclear. In particular, because at these short length scales spatial nonlocality becomes relevant and most constitutive parameters are measured near zero wavevector, these susceptibilities quickly become inapplicable for describing quantum optics at the extreme atomic scale. Simultaneously, driving nanoplasmonic systems out of equilibrium *via* laser-induced excitation introduces an ultrafast timescale to the problem. Understanding the dynamics of nonequilibrium electrons in materials is critical for a wide array of applications ranging from photovoltaic or photochemical energy conversion devices to nanoscale transistors for computing. However, several competing effects in such ultrafast experiments are obscured in typical empirical analyses. An unambiguous resolution of the experimental signatures of hot carrier dynamics using a predictive *ab initio* theory would therefore be highly desirable. This challenge naturally grows when the initially excited metal is interfaced with another material, for instance a semiconductor, bringing nanoscale energy transfer to the ultrafast regime. As we briefly introduce in this Perspective, the field has made tremendous progress toward such an *ab initio* understanding in modeling and interpreting nonequilibrium plasmonics as well as introducing first-principles calculations to unravel plasmonic chemistry at the atomic scale.

Decay of surface plasmons to excited "hot" carriers is a direction that has attracted considerable interest from the nanoscience community.[4, 8-11] A theoretical understanding of plasmon decay processes and the underlying microscopic mechanisms has enabled a spectrum of recent demonstrations in plasmonic nanochemistry. Early reports of first-principles calculations that describe all of the significant microscopic mechanisms underlying surface plasmon decay and predict the initial excited carrier distributions resulting from decay appeared in 2014. In particular, the first *ab initio* calculations of phonon-assisted optical excitations in metals were critical to bridging the frequency range between resistive losses at low frequencies and direct interband transitions at high frequencies. Similarly, calculations of energy-dependent lifetimes and mean free paths of hot carriers, accounting for electron–electron and electron–phonon scattering, offered critical insight toward the transport of plasmonically generated carriers at the nanoscale. These predictions widened the impact of nonequilibrium plasmonics to a variety of fields in which observation or exploitation of plasmonically excited hot carriers is important, including photodetection, photovoltaics, chemical transformations, and spectroscopy.

In the field of nonequilibrium plasmonics, theoretical approaches can readily account for the time dependence of the energy distribution of carriers due to electron–phonon and electron–electron scattering as well as the optical response corresponding to direct and phonon-assisted transitions, all including detailed electronic-structure effects. This finding critically results in quantitative agreement with both the spectral and temporal features of the transient-absorption measurements, a key achievement in this field. In addition, recently generalized calculation methodologies enable researchers to avoid effective electron temperature approximations, thereby retaining microscopic details and naturally describing nonthermal, thermalizing, and thermalized electrons, as we discuss next.

When a plasmonic mode is excited in a conductive nanostructure, all the energy that is not radiated, transferred through near-field interaction, or shared through charge transfer events, eventually ends up contributing to heating up the nanostructure and dissipating into its environment as heat (see **Figure 1a**). Although deploying plasmonic NPs as nanoheaters is useful for a variety of applications, including thermal photocatalysis—as described in detail in one of the next sections—this approach lacks some of the properties that make plasmonic catalysis most promising,[3] which can instead occur through charge carrier injection. This context has justified an increased interest in probing the internal descriptions of the out-of-equilibrium electronic states in plasmonic systems, to then find strategies for optimizing the excitation and extraction of these carriers. To this end, a host of modeling approaches have been employed to address the key aspects of carrier excitation, relaxation, and transfer, from atomistic *ab initio* models to hybrids of classical and quantum mechanical frameworks. Collectively, they have significantly deepened our understanding of the internal carrier dynamics in plasmonic systems.

Critically for photocatalytic applications, the optically driven excitation of carriers in a plasmonic nanostructure is constrained by the requirements of momentum conservation. Both phonon and surface scattering can enable intraband electronic transitions to high-energy states, susceptible to escaping the metal, with the latter mechanism having the additional advantage of promoting transitions near the metal–environment interfaces. As a result, the probability of successful charge carrier transfer to a molecular adsorbate is increased.[12] It has been shown that this process dominates in systems with small features,[13] pointing to the advantage of using small plasmonic NPs for charge injection. These results are in line with previous models of plasmon dephasing in small metal NPs,[14] and are consistent with recent experimental findings on size-dependent efficiency of intraband photoluminescence in Au nanorods.[15] Importantly, even though in larger plasmonic systems the excited populations are dominated by lower-energy carriers, the surface-mediated mechanism remains relevant, and we can exploit it to enhance the excitation of high-energy carriers by employing geometries with hot spots or high surface curvature, such as octopods and nanostars.[16-18]

There are, however, fundamental limits to the overall efficiencies achievable in using plasmonic hot electrons for practical applications, introduced by considerations of detailed balance.[12, 19] These limits stem from: **(1)** the fast typical relaxation times of excited carriers with large kinetic energies, arising from electron–electron scattering;[20] this process restricts their mean free paths, limiting the system's volume that can effectively generate injectable carriers; **(2)** the relatively low rates of excitation for carriers with high energies due to the limitations imposed by conservation of momentum;[12, 21] and **(3)** constraints on the magnitude and direction of the carriers' momentum required to avoid reflection at the interface.[12, 22] **Figure 1b** illustrates these factors. These constraints are fundamental and impose hard limits on the efficiency of carrier injection for a given plasmonic nanostructure and its interface with the environment.[23] However, despite these fundamental limitations, some parallel approaches for photocatalysis using plasmonic hot carriers have been introduced in recent years, reinvigorating research in the topic. By careful system design[24-29] or by the exploitation of phenomena such as strong intra- and inter-particle hot spots,[16, 30, 31] the community is constantly trying to push the limits of the experimentally measured efficiencies. The comparatively underexplored channel of direct charge promotion between metal and adsorbates also plays a relevant role,[32, 33] as an interfacial process that coexists with the indirect injection channel considered above.

Notwithstanding the number of largely compatible theoretical models for the excited populations of electrons in plasmonic materials, some gaps and differences remain, so we do not yet have a complete theoretical picture enabling the effective nanoengineering of systems for hot carrier injection. For instance, investigations of the momentum distribution of plasmonic excited electrons aimed at understanding its injection probability are rare.[34] Moreover, this aspect would benefit from being modeled in specific nanostructures, beyond bulk crystals, as well as from combining it with a detailed picture of spatial excitation and diffusion.[16, 35] In addition, the shape of the energy profile of carriers initially excited by plasmon decay is a relevant point for which there is not complete agreement among theoretical approaches. Although a typical physical picture results in a distribution of intraband excited carriers extending to a relatively flat distribution,[13, 34, 36] several groups have discussed the relevance of optical and plasmonic-relaxation processes exciting large numbers of lower energy, nonthermalized electrons in addition to high-energy (hot) electrons.[12, 21, 37, 38] **Figure 1c** shows the characteristic shapes for nonthermalized carriers under both perspectives, as well as thermalized distributions, which are similar in both cases. This figure panel assumes pulsed illumination, for clarity, but this difference is also relevant in the steady state because the fraction of the absorbed power utilized to excite high-energy carriers is different under each of the two frameworks. Another aspect that requires additional investigation from the theory-side is the effect of interfaces in hot carrier dynamics, particularly in their injection.

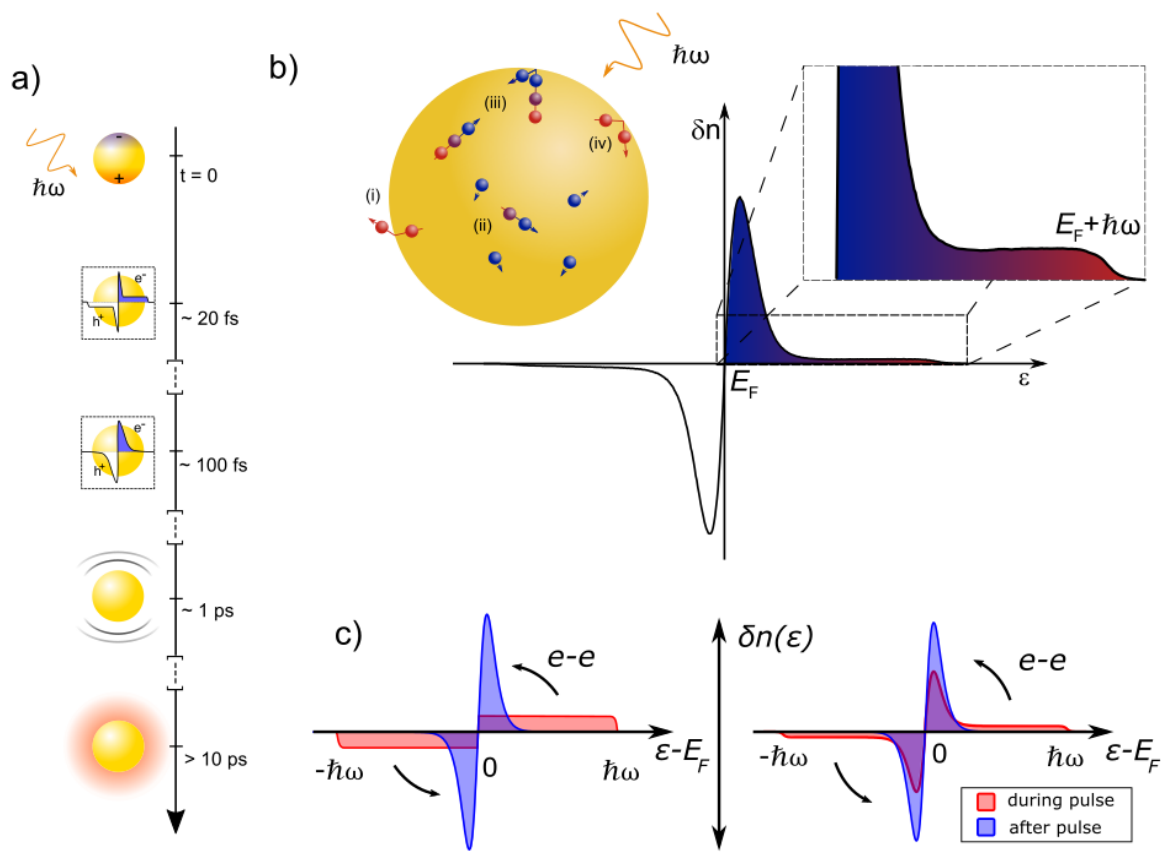

**Figure 1. a)** Schematic of the relevant ultrafast and fast timescales of different steps involved in the internal relaxation of a collective plasmonic mode in a nanocrystal after a pulsed excitation. The initial out-of-equilibrium carrier distribution excited by the plasmon thermalizes in the scale of ~100 fs, to then share their energy with the lattice and ultimately with its environment as heat. Adapted with permission from ref 39. Copyright 2019 Elsevier. **b)** Schematic of a metal nanoparticle (NP) under illumination, depicting sample trajectories of individual electrons, color-coded by their kinetic energies. The indirect injection of plasmonic excited carriers **(i)** is limited in several ways. The carriers need to have sufficient kinetic energy to surpass any potential barrier separating metal and environment, and the majority of excited plasmonic carriers will not have sufficient energy to leave the system **(ii)**. The schematic plot depicts a typical steady-state population of excited intraband carriers for a small NP, over those at equilibrium with the lattice temperature, with small overall numbers of high-energy hot electrons. These carriers have very short lifetimes, limiting their capacity to reach the surface before losing a fraction of their kinetic energy, as it has been discussed in the context of Au, Ag, Cu and Al plasmonics **(iii)**. Last, conservation of momentum at the interface also constrains the directions of propagation that avoid reflection at the surface **(iv)**. **c)** Illustration of two common pictures for the initial excited carriers' profile under pulsed excitation. Adapted from ref 40. Copyright 2017 American Chemical Society.

*Ab initio* predictions for the parameters describing the dynamics of electrons out of equilibrium with the lattice as well as the optical response of these hot electrons, have enabled a complete description of ultrafast laser measurements.[41-43] In the past few years, researchers have reported qualitative differences from previous semi-empirical estimates of electron–lattice coupling due to the strong energy dependence of the real electron–phonon matrix elements, suggesting a revision of previous empirical models for electron relaxation. By presenting comprehensive temperature- and frequency-dependent predictions of the optical response of hot electrons in plasmonic metals, theoretical and computational results have opened new avenues in plasmonic catalysis. We anticipate that open questions in metal–molecule and metal–semiconductor or metal–metal hybrids in deeply nonequilibrium regimes require a similarly focused effort to capture the dynamics uniquely localized at the interface.

Although researchers have studied surface roughness and its role in relaxing the momentum requirements for carrier injection,[12, 44] our understanding of interfacial effects can still benefit from a detailed study of the electronic configuration at the boundary between specific material systems, especially when considering adsorbed molecules. Exploring these points in a manner that targets relatively large systems while including insights from atomistic *ab initio* models can be challenging due to the disparity of spatial and temporal scales relevant in a plasmonic photocatalytic system. Thus, looking forward, computational approaches exploring hot carrier excitation and injection in realistic systems can benefit from extending existing phenomenological and multiscale models,[21, 35, 45] aiming especially to extend critical insights gained by atomistic *ab initio* methods beyond density-functional theory[13, 34, 46, 47] to larger-scale systems. The integration of information from a combination of atomistic *ab initio* methods, semi-classical optical models, and molecular dynamics could simultaneously account for the different factors limiting the carrier injection pathway in plasmon-driven catalysis, and its interaction with other energy transfer mechanisms that occur concurrently.[48] This integration may be particularly relevant when studying hybrid strategies with different nanostructures and materials, including nonhomogeneous alloys or structures of diverse dimensionality.

Despite relatively low hot carrier injection efficiencies, plasmon catalysis is still of fundamental scientific and technological relevance. As we discuss in this Perspective, indirect hot electron injection is not the only relevant energy transfer mechanism.[5] Furthermore, beyond extending our understanding of energy conversion and transfer mechanisms, the field can pursue strategies that exploit several of these mechanisms simultaneously, either synergistically or contributing to several processes in a multireaction scenario.[49] Moreover, there are new approaches for photocatalysis in which plasmonic systems provide unique advantages. A recent example is the proposed use of chiral plasmonic systems in combination with photocatalysis and photothermal effects, which, with the circular polarization of light, provides an additional dimension over which to control photocatalytic reaction rates[50] and heat generation;[51] in a complementary approach, growth induced by circularly polarized light can also introduce chirality in nonchiral structures.[52]

**Thermal Effects in Plasmonic Systems**

Major challenges in plasmonic catalysis originate from the diverse physical phenomena and broad time scales involved. Whereas plasmonic relaxation and field scattering happen on a ~fs time scale and can be considered virtually instantaneous from a chemical reaction perspective, hot carrier thermalization and lattice temperature increase (through electron–electron and phonon–phonon interactions) typically span between hundreds of fs and tens of ns, as shown in **Figure 1a**. Such a wide time window broadly overlaps with the time constants of typical chemical reaction mechanisms. As such, singling out the influence of each physical phenomenon in plasmonic reactivity is a compelling task.[53-55] In the present section, we describe some important considerations in the broader field of thermoplasmonics that can impact plasmonic catalysis, and in the next section we highlight different experimental approaches used to discriminate and to quantify thermal and nonthermal effects.

Thermal transfer is the primary energy equilibration channel between a nanoparticle (NP) and the environment. Regardless of the energy conversion pathway, a portion of radiation absorbed by NPs is always dissipated as heat. As described in the previous section, small metal NPs are particularly relevant and attractive for plasmonic catalysis so let us first analyze the thermal behavior of this type of system.

For sufficiently small metal particles at their plasmonic resonance, far-field scattering is minimized and the vast majority of the electromagnetic energy interacting with the particle ends up increasing its temperature. However, under conventional, wide-field illumination geometries, single/few NPs alone can hardly be used for practical applications in thermal-related processes, as shown in **Figure 2a**. By fixing the illumination intensity and absorption efficiency (the absorption cross-section divided by the geometric one), the temperature increases proportionally to the radius of the NP. Small particles will therefore give rise to limited, although localized, temperature increases. For example, to increase by 1 K the temperature of a 15 nm Ag nanosphere with an absorption efficiency close to 15—a relatively large value, reachable close to the particle localized plasmon resonance at a wavelength of ~390 nm—would require an input intensity of more than 20 MW/m$^2$. This large optical intensity corresponds to ~20,000 times the full spectrum solar irradiation and is significantly larger than the optical intensity used in conventional ensemble photochemical setups using wide-field illumination. For this reason, single-particle photothermal studies typically use focused lasers that combine wavelength selectivity with large intensities, to achieve significant absorption efficiencies and high temperatures. Single-particle photothermal approaches have been demonstrated for nanomaterials synthesis and could be useful, for example, for advancing nanothermolithography techniques, designing versatile optothermal nanophotonics devices, or conducting mechanistic studies in plasmonic catalysis.[56-60]

Plasmonic NPs under mild illumination intensities represent a valuable platform for large-scale photothermal driven processes when considered in ensemble. Analogous to the distance dependence of the electric potential from a charged sphere, temperature drops relatively weakly as ~1/r from the surface of a heated NP.[61] When multiple NPs are irradiated within the same system, the temperature contributions from all heat sources must be added[62] and the overall temperature of an ensemble can reach large, though typically homogeneous, values. **Figure 2a** shows simulations of the radically different temperature profiles emerging when either one single or nine differently separated NPs are irradiated at the same intensity. The resulting temperature profiles exhibit localized or uniform temperatures, depending on the total number of involved NPs and their relative distances. Baffou *et al.* have shown how such thermal patterns can be estimated for regular arrays of NPs and how to evaluate the homogeneity of the resulting temperature maps.[62] Three-dimensional (3D) or more disordered systems of NPs, which are commonly used in plasmonic catalysis, can be numerically described employing Monte Carlo-based approaches.[63, 64]

Nanoparticle-based steam generation[65] and nanophotonics-enabled solar desalination[66] have been demonstrated on the basis of cumulative photothermal effects in NP ensembles. In such cases, one major advantage of using nanostructures as heating sources is the small amount of material required to achieve efficient light-to-heat energy conversion. Following this concept, **Figure 2b** shows the minimum required thickness of a homogeneous layer absorbing 65%, 90%, and 99% of the solar spectrum, depending on the material average absorption coefficient. The cases of typical solar absorbers, such as Pyromark,[67] carbon black NPs[66] and plasmonic metasurfaces,[68] are highlighted. Due to the strong electromagnetic energy confinement effect, customized plasmonic-based layers typically need tens to hundreds of nanometers to absorb most of the solar spectrum, holding promise for large-scale, ultrathin photothermal devices. Efficient, thin light-absorbing surfaces also play a key role in flow-driven thermal oscillators.[69] In such systems, a thermal process can be described as a resonant phenomenon and enhanced at its resonant condition. Researchers have increased the efficiency of thermal water distillation, an Arrhenius-like temperature-dependent process, by ~500% by applying this resonant heat-transfer mechanism.[70] Plasmonic metasurfaces, by acting as quasi-two-dimensional (2D) heat dissipators, squeeze light-to-heat conversion into nanometric-sized layers and hold promise for enabling nanophotonics-driven ultrathin resonant thermochemical reactors.

Another property of plasmonic particles that deserves attention for thermal applications is their fast thermalization dynamics. The typical thermalization times of a plasmonic NP scale as the square of its radius,[71] and, if fast thermalization times are sought, heating timescales benefit from the small size of nanostructures. In **Figure 2c**, we plot the heating time constant as a function of the NP radius, for both water and air environments. The insets show the temperature profiles for the case of a 15 nm Ag particle immersed in water and irradiated at ~21 MW/m$^2$ after 0.1 ns, 1 ns, and 100 ns. Interestingly, macroscopic temperature increases can be achieved in the ~ns timescale, opening opportunities for the study of ultrafast nanoscale thermochemical reactions. Indeed, temperature increases about hundreds of degrees Kelvin can be achieved in focused-light single NP and wide-field NP ensemble experiments. If particles are heated at these relatively large temperatures, the absorption efficiency itself starts to exhibit temperature dependence, inducing photothermal nonlinearities that have to be accounted for in order to make accurate thermal predictions.[72] This optical absorption-temperature dependence can be engineered, for example, to increase or to decrease heat dissipation in rod-shaped nanostructures.[73]

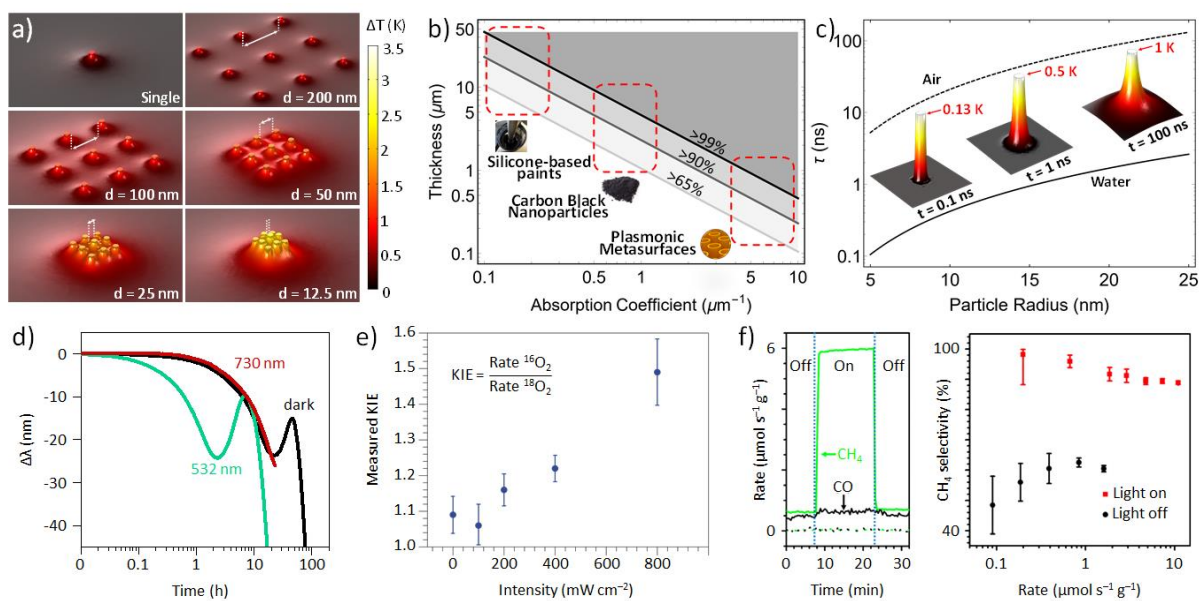

**Figure 2**. **a)** Calculated temperature profiles of a single and arrays of 9 x Ag 7.5nm radius nanoparticles (NPs) in a square lattice with lateral distance from 5 nm to 200 nm. In all cases, input light intensity is constant at $I_0 \sim 21.4 \, MW/m^2$. The absorption efficiency of each NP is assumed constant, $\eta_{abs} = 15$. Surrounding material is water. **b)** Thickness of a homogeneous light absorbing layer required to absorb 65% (light gray), 90% (dark gray), and 99% (black) of solar radiation, depending on the material absorption coefficient, averaged over the solar spectrum. Typical values for selective solar absorbers such as Pyromark 2500,[67] carbon black NPs[66] and plasmonic metasurfaces[68] are shown in red. **c)** Typical thermalization time scale, $\tau$, for single Ag NPs of different radii, in the case of water (solid) and air (dashed) surrounding medium. Insets show the temperature profiles of a 15 nm radius Ag NP after 0.1 ns, 1 ns, and 100 ns since the beginning of irradiation ($I_0 \sim 21.4\ MW/m^2$) at the particle dipolar plasmonic resonance ($\eta_{abs} = 15$). **d)** Time evolution of the longitudinal plasmon resonance of Au nanorods during the growth of a silver shell (black) in the dark, (red) under 55 mW of 730 nm laser light, and (green) under 240 mW of 532 nm laser light. The NP suspension is stirred and actively cooled to 6 °C and the total absorbed optical power under 730 nm and 532 nm irradiation is constant. Adapted from ref 53. Copyright 2020 American Chemical Society. **e)** Kinetic isotope effect (KIE) measured for the photocatalytic ethylene epoxidation (limited by the dissociation of $O_2$) on Ag nanocubes at constant reaction rate, as a function of light source intensity. Note that for the range of temperatures explored in this study, the KIE for the thermal (dark) process changes by less than 1%. Adapted with permission from ref 74. Copyright 2012 Springer Nature. **f)** Left panel: Rates of $CH_4$ (green) and CO (black) production at 623 K on plasmonic Rh nanocubes supported by $Al_2O_3$ under dark and UV illumination. Neither products are observed in a control experiment with only $Al_2O_3$ (dotted lines). Right panel: Selectivity toward $CH_4$ as a function of the overall reaction rate in dark (black circles) and under UV light (red squares). Adapted with permission from ref 75. Copyright 2017 Springer Nature.

**Differentiating Thermal and Nonthermal Effects in Plasmonic Catalysis**

As discussed in the last few sections, photocatalytic reactions on plasmonic NPs can be influenced by a variety of factors, including the generation and ejection of nonthermalized ("hot") charge carriers and the heating of the NP upon light absorption. Due to the extremely short lifetime of hot electrons and holes in plasmonic metals,[4, 76] in many cases the latter photothermal effect dominates. In this section, we address the often-subtle impact of photothermal effects on plasmonic chemistry and highlight different experimental approaches that are capable of discriminating and, in some cases, quantifying their contributions.

The rate of (photo)catalytic reactions depends exponentially on the temperature *via* the Arrhenius relation. For this reason, even small uncertainties in the determination of the temperature of the catalytic active sites can result in large over- or under-estimations of the reaction rate enhancement due to nonthermal plasmonic effects. A common method used to distinguish between thermal and nonthermal photocatalytic processes is to study the reaction rate under varying illumination powers.[74, 77-80] As several authors have pointed out, however, due to the typically limited range of accessible optical powers in practical experiments, this method alone is often insufficient to completely rule out thermal contributions to the measured rate enhancements.[81-83] As a result, additional experimental evidence for the role of nonthermal effects in plasmon-driven photocatalysis is typically necessary.

Arguably, the simplest method for determining potential thermal contributions is *via* direct measurement of the catalyst temperature during *operando* conditions.[55] Although this approach seems rather obvious, it is often experimentally challenging because of complex light absorption and scattering pathways, large temperature gradients across the samples, nontrivial mass and heat transport, the photosensitivity of thermocouples, and the uncertainty associated with infrared (IR) thermal camera measurements.[63, 64, 79, 84-87]

Processes driven by hot charge carriers are characteristically "resonant" because they strongly depend on the excitation wavelength. On the contrary, photothermal effects are dissipative and only depend on the total absorbed optical power in the NP catalysts.[88] This difference can be exploited to distinguish thermal and nonthermal activation mechanisms by comparing the reaction rate, normalized by the total absorbed optical power, under different irradiation wavelengths. Such an approach has recently been successfully applied in the study of several plasmon-driven reactions, including the photocatalytic decomposition of methylene blue on Ag nanocubes,[89] the plasmon-driven electropolymerization of aniline,[55] the electrochemical oxidation of glucose,[90] and the plasmon-driven synthesis of core–shell NPs in solution (**Figure 2d**).[53] In addition, the time differences between heat dissipation (long range) and carrier transport (short range) have been used to disentangle each contribution in plasmon-assisted electrochemical measurements on Ag electrodes.[91]

Catalytic reactions on metallic surfaces exhibit kinetic isotope effects (KIEs): conversion rates that depend on the isotopic composition of the reactant molecules (**Figure 2e**). The magnitude of these effects depends on the nature of the activation process, with electron-driven reactions having larger KIEs than phonon-driven (thermal) ones. By comparing the KIE of catalytic reactions under plasmon excitation and under dark conditions, it is possible to confirm the electronic, nonthermal nature of several plasmon-driven enhancement processes.[74, 92, 93] Other experimental approaches that have successfully been used to demonstrate the existence of nonthermal activation pathways in plasmon-driven photocatalytic reactions include ultrafast spectroscopy,[94] photoelectrochemical measurements,[55, 95] spatially resolved reactivity,[5, 16, 96] and selectivity studies (**Figure 2f**).[75, 97-99]

**Interfacial Problems and Current Solutions for Extracting the Carriers**

Although the role of hot carriers in plasmon-driven chemistry has been hotly debated,[100] the possibility of collecting them is best demonstrated by the numerous sub-bandgap photodetection devices reported in the past 10 years. Starting from the first demonstration of near-IR light detection with Au active optical antennas on Si in 2011,[9] this field has experienced rapid growth. Initially, research focused on maximizing light absorption and enabling new functionalities.[101] Subsequently, increasing effort has been devoted to understanding the generation, transport, and collection processes in hot-electron devices, and, more recently, in hot-hole devices as well. Detailed experiments combined with theoretical calculations have suggested that photodetection devices primarily operate thanks to the collection of ballistic hot carriers (**Figure 3a**).[102] Ultrafast spectroscopy measurements have also confirmed that both hot electrons and hot holes can be collected at ultrafast time scales (< 200 fs),[44, 46, 103] as shown in **Figure 3b,c**. Importantly, in the visible regime and for d-band metals, hot-electron devices are primarily limited by the energy distribution of the active charge carriers,[104] whereas hot-hole devices are hindered by their extremely short mean free paths.[36, 56] In addition, momentum matching conditions at the metal/semiconductor interface can dramatically modify the extraction efficiency. Strategies such as embedding the nano-antennas[105] and roughening the interface[106] can improve the device performance by relaxing the momentum matching conservation conditions. Recent conductive atomic force microscopy measurements have also shown that hot carriers are preferentially transferred to the semiconductor in regions of high electric field,[107] similarly to charge transfer to adsorbed molecules.[16]

Going beyond solid-state photodetectors, efficient collection of hot carriers across a metal semiconductor/interface plays an important role in plasmonic catalysis (**Figure 3a**): the long-lived charge-separated state that is established counteracts the ultrafast recombination of hot electrons and hot holes in the metal.[46, 108] Efficient hot-carrier collection also permits the realization of photoelectrodes, preventing charge build-up in the metal and avoiding the use of electron/hole scavengers in solution. Therefore, quantifying the energy distribution of the collected charge carriers as well as their collection efficiency is critical for understanding and engineering hot-carrier catalysis. Yet, many points remain unresolved. Theoretical calculations and ultrafast spectroscopy data reveal a strongly nonequilibrium energy distribution at early timescales[13, 109] (**Figure 3d**). In addition, theoretical modeling that tracks the spatiotemporal distribution of the charge-carrier population with and without hot-carrier extraction suggests that ballistic collection of high-energy carriers in the proximity of electric-field hot-spots,[16] which typically occur close to the metal/semiconductor interface, may be possible.[35] This result is in excellent agreement with the modeling of ballistic collection of high-energy electrons/holes above a Schottky barrier in solid-state devices.[36, 46, 102] However, recent experimental quantifications of the energy distribution of the hot carriers[110, 111] based on scanning microscopy show a charge-carrier population with an extremely small fraction of high-energy electrons/holes (**Figure 3e**). This result matches well with calculations of the steady state distribution of plasmonic hot carriers that do not include any ballistic harnessing of the highest energy carriers.[111] Importantly, recent ultrafast spectroscopy results suggest that the removal of highly energetic carriers of one type (*e.g.*, holes) results in a lower thermalization temperature and, hence, lower peak energy of the complementary carrier (*e.g.*, electron) that remains in the metal.[46]

Overall, thorough modeling and experimental research are critical to clarify the effects of charge separation across a metal/semiconductor interface on the hot-carrier population that remains available for photocatalysis. In this respect, as shown by recent results,[112] it is necessary to improve control over the interface properties (**Figure 3g**). Furthermore, advanced ultrafast experimental techniques are needed to probe these electronic processes at different time scales.[113, 114]

Remarkably, a major discrepancy remains between the hot-carrier-collection efficiencies estimated by ultrafast spectroscopy experiments (> 40%; as shown in refs [32, 44, 46]) and those measured in solid-state or photoelectrochemical devices (<1%) under steady-state conditions. Although interfacial excitation can enhance charge separation in metallic nanocrystals (**Figure 3f**),[32, 113] such discrepancy persists for metal/semiconductor interfaces employed in larger scale devices that are accurately described by a three-step generation and collection model (*e.g.*, Au/p-GaN). Therefore, following the example of recent spectroscopic analysis,[115] further studies are needed to understand the role of interfacial states on plasmon dephasing as well as hot carrier generation and collection. Importantly, a deeper understanding of quantization effects on the thermalization of plasmonic hot carriers[116] must be achieved to compare high-intensity ultrafast studies with low-intensity solid-state and photoelectrochemical device measurements correctly. Further, the role of nonlinearities in high-field plasmonic systems and modulation of plasmonic absorption by ballistic thermal injection remain open questions.[117, 118] Overall, resolving these apparent contradictions and advancing the understanding of hot carrier collection processes is of the greatest importance for the future of plasmonic hot carrier catalysis.

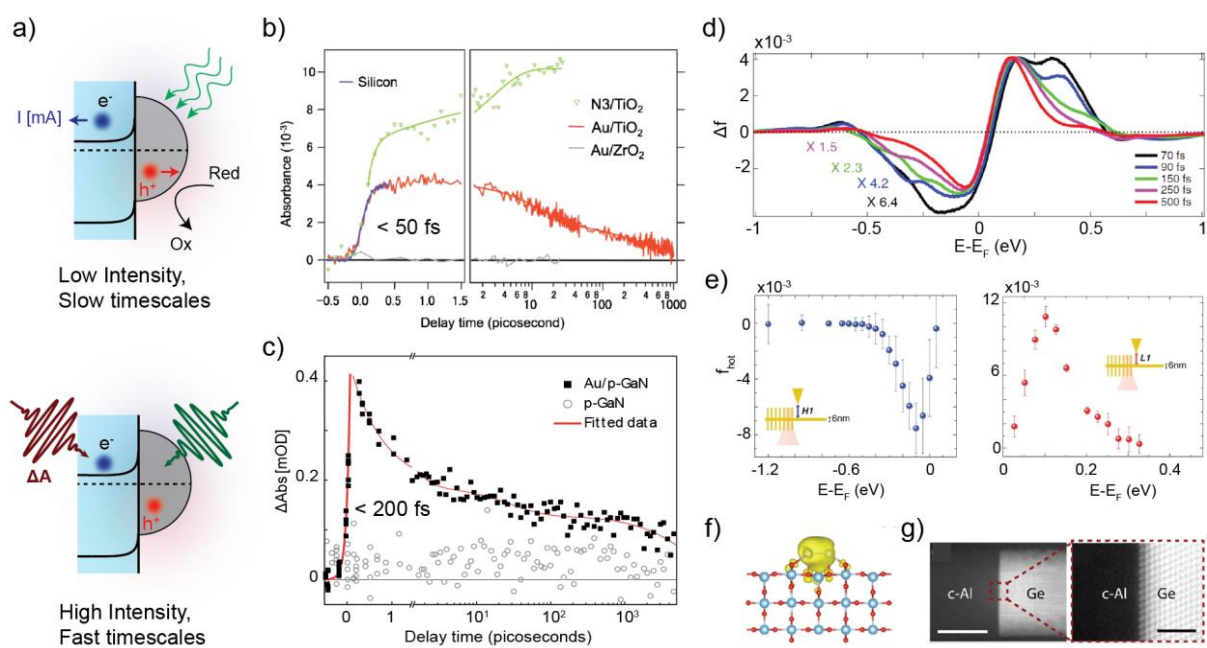

**Figure 3. a)** Schematic depiction of the hot carrier separation across a metal (grey)/n-type semiconductor (light blue) interface during steady-state photoelectrochemical measurements (top) and probed with ultrafast spectroscopy methods (bottom). **b)** Infrared ultrafast spectroscopy of hot electron injection from Au nanoparticles into n-type $TiO_2$ (red curve). A sub 50 fs injection time is estimated from the rising edge. Adapted with permission from ref [108]. Copyright 2013 Elsevier. **c)** Infrared ultrafast spectroscopy of hot-hole injection from Au into p-type GaN (black squares). A sub-200 fs injection time is estimated from the instrumented limited rising edge. Adapted with permission from ref 46. Copyright 2020 Springer Nature. **d)** Hot electron ($E-E_F > 0$) and hot hole ($E-E_F < 0$) charge-carrier distributions in an Au film as a function of time estimated from ultrafast spectroscopy data. Adapted with permission from ref [109]. Copyright 2018 Springer Nature. **e)** Hot hole (left) and hot electron (right) charge-carrier distribution in an ultrathin Au film measured with scanning tunneling microscopy under steady-state surface plasmon polariton excitation. Adapted with permission from ref [111]. Copyright 2020 American Association for the Advancement of Science. **f)** Density functional theory calculation of the charge distribution at an Ag nanocrystal/$TiO_2$ interface. Reprinted with permission from ref [113]. Copyright 2017 Springer Nature. **g)** Transmission electron micrograph showing a highly controlled aluminum/germanium interface used to study hot electron injection. Reprinted from ref [112]. Copyright 2020 American Chemical Society.

**Colloidal Structures for Plasmonic Photocatalysis**

In the same way that semiconductor–metal junctions have been proposed to increase the lifetime of plasmonic electron–hole pairs,[119] catalytic–plasmonic metal heterostructures have been postulated to merge the reactive and photon-harvesting abilities of different materials.[120-122] A group of hybrid colloids designed to act as plasmonic catalysts emerged recently, including core@shell, antenna-reactor, and hetero-dimers, to name a few. In general terms, these bimetallic systems have shown higher efficiencies compared to their monometallic counterparts. However, these methods of combining plasmonic and catalytic metals have failed to improve the efficiency of photocatalytic reactions significantly, and we are still far from having materials of industrial relevance. A closer inspection reveals that photon absorption is the first step and injection of the energy into molecules is the last step in a complex process that needs to be optimized to increase the efficiency of light (and possibly sunlight) driven plasmonic catalysis. Energy storage and distribution in hybrid systems such as those described above are a fundamental open question, as discussed in the previous sections.

Having excellent charge harvesters and injectors such as the plasmonic and catalytic examples, respectively, is the first step toward increasing the efficiency of photoreactions. However, the best possible method of combining them remains unknown, highlighting the critical role of the interface for extracting energetic carriers. **Figure 4** shows transmission electron microscopy (TEM) images of hybrid colloidal structures fabricated to work under many different mechanisms: hot-carrier separation,[123] direct injection of plasmonic hot electrons into a second material,[121] near-field-induced enhanced absorption on catalytic metals,[122] and near-field increased at the surface of the catalyst,[124] among many others.

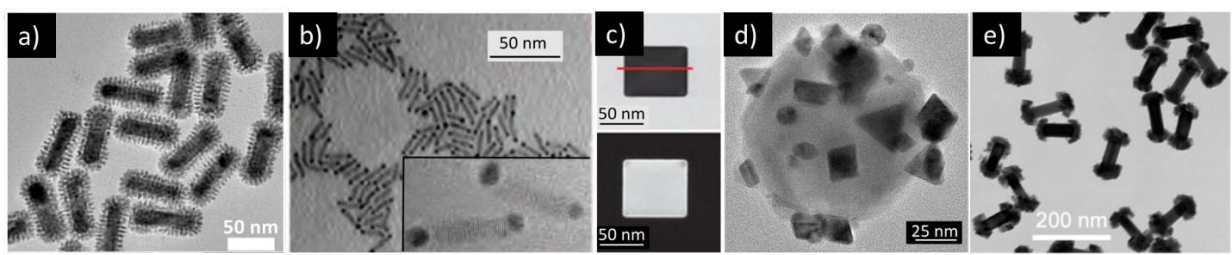

**Figure 4**. Transmission electron microscopy images of different types of hybrid colloids proposed and tested for plasmonic catalysis. **a)** Au@Pd nanorods. Reprinted from ref [125]. Copyright 2017 American Chemical Society. **b)** CdSe-Au hybrid nanorods. Reprinted with permission from ref 32. Copyright 2015 American Association for the Advancement of Science. **c)** Ag@Pt core–shell nanocubes. Reprinted with permission from ref [121]. Copyright 2017 Springer Nature. **d)** Al-Pd antenna–reactor complexes (image credit: D. Swearer/Rice University). **e)** Au-$CeO_2$ site-selective nanorods. Reprinted from ref [123]. Copyright 2019 American Chemical Society.

For metal–metal interfaces, such as those shown in **Figure 4a, c,** and **d**, the electrons tend to flow in the direction that decreases the gradient of their electrochemical potential until it vanishes. This process is known as Fermi level equilibration and it can be highly dependent on the size of the structures and the contact surface area between materials.[126] This dependence points toward a highly interface-dependent electronic structure in bimetallic colloidal systems. Also, the electrons' and holes' removal capabilities can be fairly different depending on the bimetallic system under consideration; in some systems there is only one metal exposed for both processes whereas in others there are two different metal sites for extraction. However, colloids shown in **Figure 4b** and **e**, have similar metal-semiconductor interfaces to those described in the previous section. As such, plasmon excitation and decay in each of the systems described in **Figure 4** are fundamentally different. Further understanding of these processes could help to optimize the materials, shapes, and combination methods at the nanoscale.

Recent advances in the ultrafast response of bimetallic core@shell systems, for example, point toward plasmon-enhanced dephasing and energy transfer to a second metal at the interface as the dominant plasmon-decay mechanism.[127] This energy transfer resembles the plasmon-induced interfacial charge transfer mechanism in metal–semiconductor structures.[32] This cross talk between plasmon excitation and a second dephasing material can also be extended to molecules. Molecular adsorbates can act as scattering centers, also favoring plasmon dephasing in a phenomena known as chemical interface damping (CID).[128] These phenomena are fundamentally different from those operating in antenna-reactor types of complexes, where a detailed balance of local field enhancement, scattering, and absorption determines their catalytic performance.[129] Also, as mentioned earlier, the possibility of extracting both type of carriers (electrons and holes) is remarkably different in each of the structures shown in **Figure 4**. This fact is often overlooked in colloidal plasmonic catalysis: the energetic requirements of the counter reactions as well as the composition and shape of the colloids can contribute to the overall efficiencies in these systems.[17, 130-132] Further understanding of energy management in hybrid systems could illuminate and guide the synthesis of a new generation of colloidal plasmonic catalysts.

In addition to the efficiency of colloidal plasmonic catalysts, the selectivity aspect of chemical transformations using sunlight is particularly promising for the field.[75, 89, 133, 134] The possibility of using sunlight to alter the activation barrier of certain chemical reaction pathways is undoubtedly a major step toward controlling chemical reactivity by external means.[120] In order to advance toward this goal, unraveling the reactive sites and their wavelength dependence reactivity on the surface of these hybrid plasmonic systems is of utmost importance.[135] Because the chemical bond is by nature on the atomic length scale, final control of reactive sites on colloidal structures should aim at controlling colloids at the atomic scale.[136] Some recent and elegant studies on single-atom plasmonic catalysis have shown great results at achieving atomic control on plasmon reactivity,[137, 138] and this is an emerging but highly promising area to move the field forward.

**Atomic-Scale Plasmonics for Catalysis**

Plasmonic picocavities are formed in metal–insulator–metal structures when a single metal atom protrudes out of one of the metal structures, as shown in **Figure 5a-c**. The first picocavity was realized in a nanoparticle on a mirror (NPoM) configuration, which is composed of a flat, Au surface and an Au NP separated by a molecular monolayer (see **Figure 5a–c**).[139] An earlier work showed that under continuous-wave laser irradiation (447 nm, 0.2 mW), one can controllably reshape the geometry of a nanoplasmonic structure, tuning its optical properties.[140] The laser produces a strong field enhancement, delivering energy to the surface Au atoms on the NP. Some of the surface Au atoms, especially at atomic sites with local defects where surface energy is lower, are mobilized and migrate, changing the geometry of the nanoplasmonic structure.

To produce stable and robust picocavities where only a single Au atom is mobilized and stabilized at a specific position, one needs to perform a similar but more controlled experiment. In the seminal work from Baumberg and co-workers,[139] the plasmonic system (NPoM) was cooled at cryogenic temperatures to reduce thermal effects in reconfiguring the surface. The laser pump was kept at sub-100 µW power levels and the laser irradiation wavelength was set at 633 nm to ensure that the field enhancement was in the middle of the gap, where the picocavity should be formed. The energy delivered by the laser enabled a single Au atom to be pulled down and to protrude from the NP, as shown in **Figure 5c**. The protruded atoms remained in position (*i.e.* stabilized) as long as the laser power was kept at low values. The single Au atom protruding out of the NP facet may be pulled from either a vacant atom site or from local defects formed at the junction of two or more different crystalline Au-plane orientations that exist in the NP.

Picocavities have the ability to enhance plasmon fields even further and to produce strong field gradients, as shown in **Figure 5a**. The picocavity (*i.e.* the single Au atom protrusion) does not give rise to additional resonant modes and it is not resonant itself (at least not at optical frequencies), but instead only provides an additional field enhancement super-positioned on the existing plasmon mode fields without the protrusion (**Figure 5b–c**). The additional field enhancement is confined at extremely small volumes,[141] enough to encompass a single molecule (see **Figure 5c–d**). In addition, the strong field gradients from the picocavity can illuminate only parts of a single molecule, activating specific vibrational modes of chemical bonds within a single molecule (see **Figure 5d**).[139] This effect appears as additional surface-enhanced Raman spectroscopy (SERS) lines in both Stokes and anti-Stokes Raman spectra of the molecule in the picocavity. Using density functional theory (DFT) numerical calculations, the authors identified the relative position of the molecule with respect to the picocavity for the measured extra SERS lines that emerged (see **Figure 5d**). Since then, a few more tip-based techniques with atomic light confinement have been reported, achieving Å resolution for imaging the normal modes and intramolecular charges/currents in single molecules, either using the Raman signal emerging from specific chemical bonds,[142, 143] or photoluminescent measurements at a submolecular level.[144] A recent theoretical work combined DFT and classical calculations to map the Raman behavior of single molecules in strongly inhomogeneous fields, and demonstrated that picocavities are capable of probing and revealing "intramolecular features of a single molecule in real space".[145] Hence, picocavities provide unprecedented access to specific chemical bonds within a single molecule, opening pathways to modulate and to control optically specific intramolecular bonds and site-selective chemistry.[141]

Although thus far picocavities have been used to access vibrational modes at the submolecular level, they have the potential to offer extraordinary control over catalytic reactions at the single-molecule level as well. Recent advances in following chemical transformations at the single-molecule level using enhanced Raman spectroscopy bring us one step closer to this goal.[146-151] However, there are open questions on many different fronts related to the formation of picocavities, how hot carriers are generated at picocavities, and how a picocavity can affect catalytic reactions at the single-molecule level.

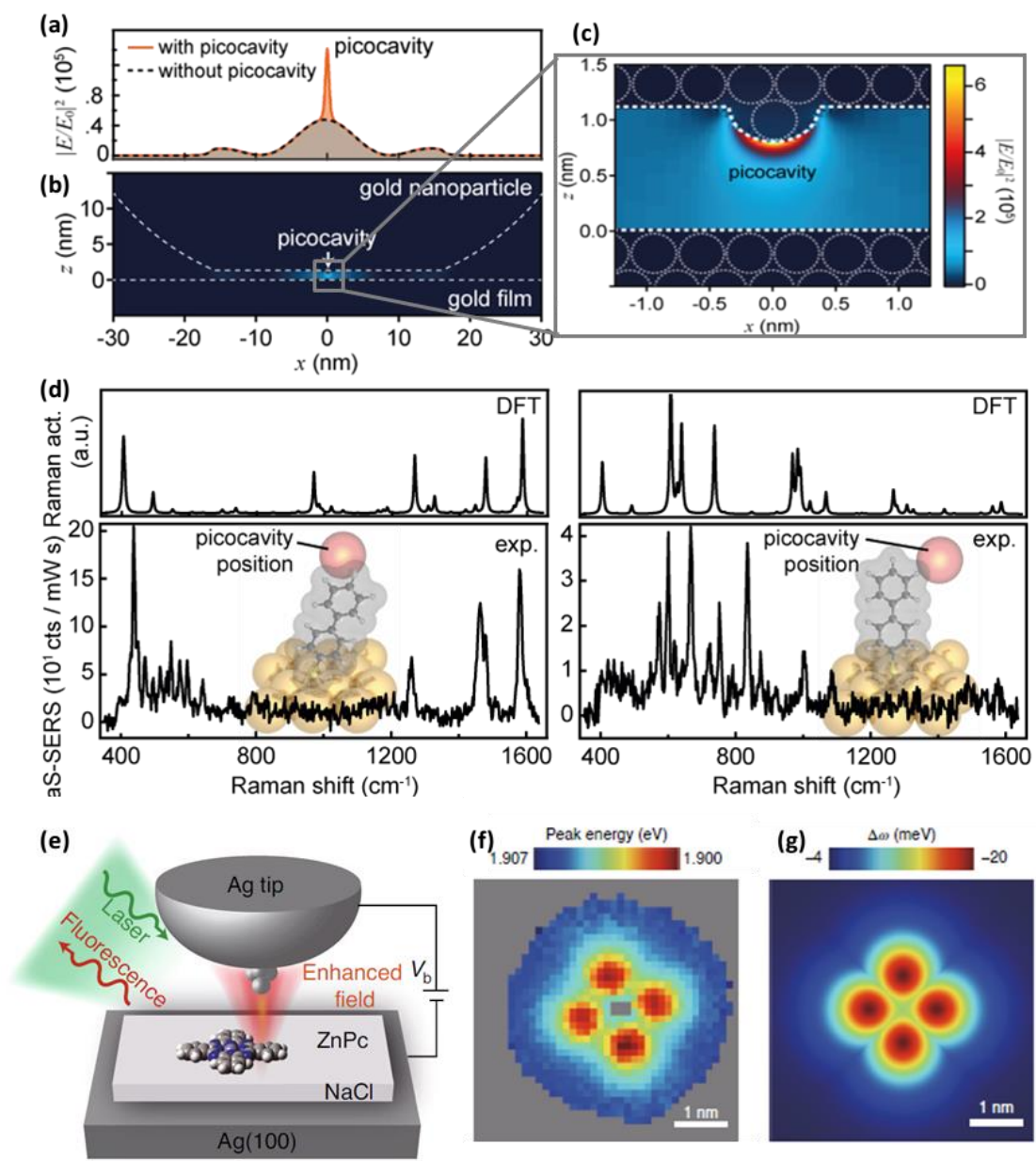

**Figure 5.** Plasmonic picocavities. **a)** The picocavity creates an additional field enhancement localized around the atomic protrusion (full orange line) compared to just the plasmonic nanocavity (dashed black line). **b)** Diagrammatic representation of the picocavity relative to the nanoparticle and nanocavity size, with **c)** showing the field gradients around the atomic protrusion. The dashed circles indicate the Au atoms. Adapted with permission from ref [139]. Copyright 2016 American Association for the Advancement of Science. **d)** The relative positioning of the atomic protrusion (picocavity) with respect to the molecule leads to the emergence of additional Raman lines for both density functional theory predictions and experimental measurements. Adapted with permission from ref [139]. Copyright 2016 American Association for the Advancement of Science. **e)** Schematic representation of the experimental set-up for an Ag tip-apex with an atomic protrusion that produces confined electromagnetic fields to image a single ZnPc molecule. Adapted with permission from ref [144]. Copyright 2020 Springer Nature. Peak energy maps obtained **(f)** experimentally and **(g)** numerically. Adapted from ref 17. Copyright 2020 American Chemical Society.

It is currently believed that picocavities are formed at specific atomic sites where localized deformities between different crystalline lattices exist. We also know that the chemical reactivity of chemisorbed molecules depends on the molecule orientation with respect to the surface.[124, 152] Still, questions remain. Can a picocavity trigger a catalytic reaction on a single molecule? Would the same molecule exhibit different catalytic properties if placed in a different position relative to the picocavity? At the picocavity level, where a single Au atom is involved, quantum effects are expected to be more prominent, with the Au atom protrusion electron orbitals overlapping with the single-molecule orbitals in a different manner than a flat metal surface would. These quantum effects are expected to affect the hot-carrier transfer process, but it is not yet clear how hot carriers emerge from picocavities. Can picocavities increase the injection efficiency of ballistic electrons because there is less probability for collisions to occur? What roles do defects and roughness play in the plasmonic reactions that have been explored so far? Does the injection of picocavity hot carriers into a single molecule depend on the positioning of the molecule with respect to the picocavity? Can the strong field gradient of a picocavity modify specific chemical bonds within a molecule such that more efficient absorption of hot carriers occurs, or enable a catalytic reaction that otherwise would not occur? To the best of our knowledge, no experiments have been reported with picocavity hot carriers that are able to measure or to distinguish such phenomena. In addition, theoretical modeling of picocavity hot carriers is in its infancy and extremely challenging because it requires a combination of quantum and classical descriptions that are usually applicable at different length scales.

**Quantum Electrodynamical Control of Chemical Dynamics and Reaction Pathways**

A complementary regime of molecule–material coupling relevant to nanophotonic catalysis is the "strong-coupling" regime. Driven by ever-increasing capabilities in the fabrication of nanostructured systems, novel states of matter with hybrid and strongly correlated light–matter characteristics have been created.[153] In this regime, where light and matter intersect on the same quantized footing, quasi-particles that have both light and matter characteristics, polaritons, are formed. Research has driven the effective strength of the light–matter interaction to the strong coupling regime, and opened new avenues for quantum chemistry and catalysis paradigms. Traditionally, the quantum nature of light has been the focus of the established field of quantum optics. In recent years, however, various other research fields have also leveraged the quantum nature of light to drive molecular and material transformations. By interfacing these different fields, the strong coupling of light and matter has been demonstrated to modify and to improve various properties of technological interest dramatically. Specifically, experimentalists have demonstrated light–molecule coupling in the strong-coupling quantum optical regime. In the strong-coupling regime, the original constituents of the system lose their individual identity and hybrid quasi-particles with novel characteristics are formed. These hybridized states with mixed light–matter characteristics can dramatically change the chemical landscape.[154] Experiments in strong-coupling chemistry have presented the intriguing possibility of changing chemical reactions or controlling their selectivity.[155-159]

Combining approaches from diverse fields presents an opportunity to create a predictive theoretical and computational approach to describe cavity-correlated chemical dynamics from first principles.[154,160] Many key theoretical challenges remain. In this novel regime, the well-established theory concepts of quantum chemistry, nanophotonics, and quantum optics are no longer individually sufficient. Traditional methods from quantum chemistry such as the Born—Oppenheimer approximation, Hartree-Fock theory, coupled-cluster theory, or DFT describe the quantum mechanical nature of the states in matter, but they do not account for the quantized nature of the electromagnetic field. On the other hand, methods from quantum optics typically explore the quantum electromagnetic field in considerable detail, but the molecule is described *via* simplified models. Both over-simplifications, in quantum chemistry and quantum optics, are far from a realistic *ab initio* description of the relevant chemical dynamics, where many atoms and molecules (each with electronic and nuclear degrees of freedom) are coupled strongly to the vacuum electromagnetic field. In this few-photon limit, the quantized nature of the electromagnetic field is a key aspect that has to be considered to describe fluctuations, spontaneous emission, polariton-bound states, or thermalization correctly in tandem with a description of the quantum chemical behavior of the material–metal or metal–molecule hybrid. From a theoretical perspective, it is particularly exciting to note the intersection between plasmonic nanochemistry and the work in strong-coupling chemistry. We anticipate several advances and breakthroughs in the field at this vibrant intersection.

**Perspective**

To offer a future outlook, we provide our overview of the current bottlenecks and future challenges in the field. **Table 1** summarizes a series of topics and ideas that we believe will surface in the field in the next couple of years. We envision remarkable progress in the field of plasmonic catalysis if some of these challenges can be tackled in the near future.

Theory

*Bottlenecks:*

- **Scale disparity.** The disparity between the relevant spatial and temporal scales of different mechanisms in plasmonic catalysis challenges our ability to create computational models with a general, explicit description of plasmon catalysis.

- **Molecular-plasmonic dynamics.** Similarly, the differences in characteristic scales between energy-transfer mechanisms and molecular dynamics represent an obstacle for including both in a holistic computational description of the photocatalytic process.

- Comprehensive theoretical approaches to model the **excited states** involved in photocatalytic reactions on plasmonic surfaces.

*Challenges:*

- **Efficiency estimates.** Provide detailed estimates for maximum attainable efficiencies for the different channels of energy transfer, both for specific NP reactors and in different optical regimes to set clear goals for experimentalists and to create a framework for seeking alternative photocatalytic strategies that circumvent some premises of the models.

- **Multiscale methods.** Develop robust multiscale methods targeted at answering specific questions about mechanism interactions and serving as guiding tools for detailed nanoengineering of efficient plasmonic reactors.

Mechanisms

*Bottlenecks:*

- **Reproducibility and control.** Strong dependence of dominant activation mechanism on experimental parameters such as photocatalyst (size, shape, composition, surface chemistry), reactants (concentration, phase, flow rate, pressure), illumination (wavelength, intensity, flux), and reactor geometry (2D, 3D, colloidal suspension, gas phase).

- **Activation pathways**. Claims of nonthermal activation pathways should be corroborated by a description of which nonthermal mechanism is active (hot electrons, hot holes, near-fields) and a discussion of its limitations (generation, extraction efficiencies, average energy of hot charge carriers, intensity of near-field enhancements, etc.).

*Challenges:*

- **Focus on model systems.** 1) Model catalytic reactions with simple, well-established mechanisms, for which we can exclude side-reactions, and for which reaction rates are easily and quantitatively assessed (*e.g.*, *via* absorption spectroscopy). 2) Model samples and reactor geometries for which we can quantitatively measure or predict the optical and thermal properties under *operando* conditions. 3) Model reactant molecules for which we already have independent information on their electronic and vibrational states in homogeneous media and when adsorbed to a metallic surface.

- Experiments should carefully explore the typically **multidimensional parameter space** of photochemical plasmonic reactions. It would be desirable to vary one, and only one, of the following parameters at the time: 1) excitation wavelength, total absorbed optical power, and sample temperature; 2) NP size, composition, and surface chemistry; and 3) reactant mix, plasmonic photocatalyst, and illumination geometry.

Thermal Effects

*Bottlenecks:*

- **Size or intensity for achieving high temperatures**. To achieve high temperatures – that can assist photothermal reactions – it is possible to use either large plasmonic systems or high illumination intensities. Wide-field setups and mild irradiation intensities are effective for heating only when using large ensembles of plasmonic nanoparticles, where collective thermal effects are at play. Small arrays of plasmonic NPs reach sizable temperatures only when interacting with very high intensity light sources.

- **Thermal confinement**. To localize photothermal reactions in proximity to plasmonic NPs, either single/few particles should be irradiated or the particles need to be distanced from each other. If high temperatures are sought on a large scale, temperature localization close to the particles is lost because collective thermal effects tend to homogenize the overall thermal map. To recover the temperature selectivity near the particles, the distance among them can be increased but higher irradiation intensities are required to maintain the same peak temperatures.

*Challenges:*

- **Small is fast**. Ultrafast studies (~ns) of thermal processes at the single (or few) particle level. Temperature localization is enabled by single/few particle systems and, if high intensity sources are available, photothermal studies at the single particle level can exploit localized temperatures and large nanoscale temperature gradients. Additionally, small particles thermalize fast and their fast temperature increase can be utilized to monitor nanoparticle accelerated thermo-chemical reactions at the nanosecond timescale.

- **Large-scale metasurfaces.** Exploit large scale and highly absorptive plasmonic metasurfaces as large power density sources to sustain nanophotonics-based thermo-driven chemical processes. While large-scale plasmonic systems typically lead to homogeneous temperature patterns, plasmonic metasurfaces can be optimized to dissipate the incident light in much thinner layers than conventional absorbers. In this context, easily realizable and cost effective plasmonic metasurfaces can lead to large-scale ultrathin light-to-heat converters which potential applications in compact thermal reactors and possibly adaptable to mechanically flexible substrates.
- **Temperature feedback.** Properly design high-temperature plasmonic systems by taking into account the mutual influence between optical and thermal responses. The optical response of plasmonic system at high temperatures is modified and the light-to-heat conversion efficiency depends on the temperature itself in a non-trivial way. Future plasmonic-based thermal reactors will exploit the nonlinear photothermal phenomena for system optimization at different temperature regimes.

Applications

*Bottlenecks:*

- Lack of quantitative feasibility studies on plasmonic photocatalytic processes at relevant **industrial levels**. Our prediction is that such studies would likely indicate the presence of two major bottlenecks in the industrial application of plasmonic catalysis: the small penetration depth of visible light through plasmonic photocatalysts (in bulk-reactors) and the high capital costs associated with **photocatalytic reactors**.

*Challenges:*

- Focus on **high-added-value products and two-dimensional flow photochemistry**. In addition, given the potentially limited scope for plasmonic catalysts in large-scale industrial reactors, apply the lessons learned on hot charge carrier generation and ejection, near-field manipulation, and localized and collective photothermal effects to other fields of applications, such as photothermal therapy, light-driven drug delivery, single (bio)molecule sensing, and optoelectronic devices.
- Exploit robust nanofabrication techniques for **larger scale nanophotonics**-based plasmon- and thermo-driven chemical processes.
- Develop **highly stable nanomaterials** that can compete with current technology and thermal catalysts. Examine sintering and poisoning effects in plasmonic catalysts.

Materials

*Bottlenecks:*

- **Collection efficiency.** Understand the differences in hot carrier collection at ultrafast time scales and in steady state to reconcile current experimental observations.
- **Thermalization dynamics.** Understand the effects of charge-carrier removal as well as quantization onto the thermalization dynamics of carriers in the metal.

- **Extraction mechanisms.** Quantitatively distinguish and compare interfacial excitation and three-step injection.
- Correlate **composition and morphology** with carrier-extraction dynamics in bimetallic and hybrid colloids.

*Challenges:*

- **Device efficiency.** Improve efficiency of steady-state devices beyond 5%.
- **Targeted charge transfer.** Design the most desirable hot carrier energy distribution tailored for collection at the desired energies.
- **Sustainable plasmonics.** Toward abundant plasmonic materials for accessing the excited state in molecular targets.

Atomic Plasmonic

*Bottlenecks*:

- Atomic-scale **characterization** of picocavities by independent methods.
- Determining how picocavities are **formed**.
- **Dynamics** of picocavities at room temperature.

*Challenges*:

- **Experimental realization** of catalytic reactions with atomic picocavities at ambient conditions.
- **Controlling selectivity** in photochemical and photocatalytic reactions in a picocavity by tailoring the excitation of different reactive bonds/pathways within a molecule.
- **Theoretical modeling** of the processes involved for catalytic reactions in picocavities and at the single-molecule level.

| **Current bottlenecks** | **Topic** | **Future challenges** |
|---|---|---|
| • Scale disparity<br>• Molecular-plasmonic dynamics<br>• Excited states in photocatalysis | **Theory** 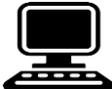 | • Efficiency estimates<br>• Multiscale methods |
| • Reproducibility and control<br>• Activation pathways | **Mechanisms** 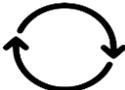 | • Focus on model systems<br>• Explore multidimensional parameter space |
| • Size or intensity for heating<br>• Thermal confinement | **Temperature** 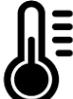 | • Small is fast<br>• Large-scale metasurfaces<br>• Temperature feedback |

| | Applications 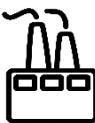 | |
|---|---|---|
| • Feasibility of plasmonic catalysis in practical applications<br>• Industrial level trials<br>• Simple reactions are usually linked to low-value products<br>• Scale-up photocatalytic hardware | | • High added value chemicals<br>• 2D flow photochemistry<br>• Cheap / large scale production<br>• Stable nanomaterials<br>• Knowledge transfer to other fields |
| • Collection efficiency<br>• Thermalization dynamics<br>• Extraction mechanisms<br>• Composition and morphology in hybrid colloids. | Materials 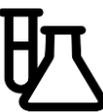 | • Device efficiency >5%<br>• Targeted charge transfer<br>• Sustainable plasmonics |
| • Atomic scale characterization<br>• Formation of picocavities<br>• Dynamics of atomic vacancies | Atomic level 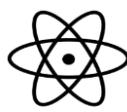 | • Experimental realization<br>• Controlling selectivity<br>• Theoretical modeling for hot-carriers in picocavities |

**Table 1.** Our vision on current bottlenecks and future opportunities for the field.

**Acknowledgments**

E.C. acknowledges funding and support from the Deutsche Forschungsgemeinschaft (DFG, German Research Foundation) under Germany´s Excellence Strategy – EXC 2089/1 – 390776260, the Bavarian program Solar Energies Go Hybrid (SolTech), the Center for NanoScience (CeNS) and the European Commission through the ERC Starting Grant CATALIGHT (802989). L.V.B. acknowledges support from the Xunta de Galicia (Centro singular de investigación de Galicia accreditation 2019-2022) and the European Union (European Regional Development Fund - ERDF). A.A. acknowledges support from the National Science Foundation under Grant No. IIP-1941227. A.B. acknowledges support from the Dutch Research Council (NWO) *via* the Vidi award 680-47-550. A.D. acknowledges support from a Royal Society University Research Fellowship URF\R1\180097 and Royal Society Research Fellows Enhancement Award RGF\EA\181038. P.N. acknowledges support from the DOE Photonics at Thermodynamic Limits Energy Frontier Research Center under Grant No. DE-SC0019140 and as a Moore Inventor Fellow through Grant GBMF8048 from the Gordon and Betty Moore Foundation.

**TOC Figure:**

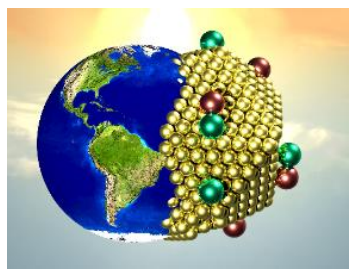

**References**

1. Buntin, S. A.; Richter, L. J.; Cavanagh, R. R.; King, D. S., Optically Driven Surface Reactions: Evidence for the Role of Hot Electrons. *Physical Review Letters* **1988,** *61* (11), 1321-1324.
2. Bonn, M.; Funk, S.; Hess, C.; Denzler, D. N.; Stampfl, C.; Scheffler, M.; Wolf, M.; Ertl, G., Phonon- Versus Electron-Mediated Desorption and Oxidation of CO on Ru(0001). *Science* **1999,** *285* (5430), 1042-1045.
3. Linic, S.; Aslam, U.; Boerigter, C.; Morabito, M., Photochemical transformations on plasmonic metal nanoparticles. *Nature Materials* **2015,** *14* (6), 567-576.
4. Brongersma, M. L.; Halas, N. J.; Nordlander, P., Plasmon-induced hot carrier science and technology. *Nature Nanotechnology* **2015,** *10* (1), 25-34.
5. Gargiulo, J.; Berté, R.; Li, Y.; Maier, S. A.; Cortés, E., From Optical to Chemical Hot Spots in Plasmonics. *Accounts of Chemical Research* **2019,** *52* (9), 2525-2535.
6. Proppe, A. H.; Li, Y. C.; Aspuru-Guzik, A.; Berlinguette, C. P.; Chang, C. J.; Cogdell, R.; Doyle, A. G.; Flick, J.; Gabor, N. M.; van Grondelle, R.; Hammes-Schiffer, S.; Jaffer, S. A.; Kelley, S. O.; Leclerc, M.; Leo, K.; Mallouk, T. E.; Narang, P.; Schlau-Cohen, G. S.; Scholes, G. D.; Vojvodic, A.; Yam, V. W.-W.; Yang, J. Y.; Sargent, E. H., Bioinspiration in light harvesting and catalysis. *Nature Reviews Materials* **2020**.
7. Zhang, Y.; He, S.; Guo, W.; Hu, Y.; Huang, J.; Mulcahy, J. R.; Wei, W. D., Surface-Plasmon-Driven Hot Electron Photochemistry. *Chemical Reviews* **2018,** *118* (6), 2927-2954.
8. Aslam, U.; Rao, V. G.; Chavez, S.; Linic, S., Catalytic conversion of solar to chemical energy on plasmonic metal nanostructures. *Nature Catalysis* **2018,** *1* (9), 656-665.
9. Knight, M. W.; Sobhani, H.; Nordlander, P.; Halas, N. J., Photodetection with Active Optical Antennas. *Science* **2011,** *332* (6030), 702-704.
10. Schirato, A.; Maiuri, M.; Toma, A.; Fugattini, S.; Proietti Zaccaria, R.; Laporta, P.; Nordlander, P.; Cerullo, G.; Alabastri, A.; Della Valle, G., Transient optical symmetry breaking for ultrafast broadband dichroism in plasmonic metasurfaces. *Nature Photonics* **2020**.
11. Halas, N. J., Spiers Memorial Lecture Introductory lecture: Hot-electron science and microscopic processes in plasmonics and catalysis. *Faraday Discussions* **2019,** *214* (0), 13-33.
12. Khurgin, J. B., Fundamental limits of hot carrier injection from metal in nanoplasmonics. *Nanophotonics* **2020,** *9* (2), 453.
13. Brown, A. M.; Sundararaman, R.; Narang, P.; Goddard, W. A.; Atwater, H. A., Nonradiative Plasmon Decay and Hot Carrier Dynamics: Effects of Phonons, Surfaces, and Geometry. *ACS Nano* **2016,** *10* (1), 957-966.
14. Genzel, L.; Martin, T. P.; Kreibig, U., Dielectric function and plasma resonances of small metal particles. *Zeitschrift für Physik B Condensed Matter* **1975,** *21* (4), 339-346.
15. Ostovar, B.; Cai, Y.-Y.; Tauzin, L. J.; Lee, S. A.; Ahmadivand, A.; Zhang, R.; Nordlander, P.; Link, S., Increased Intraband Transitions in Smaller Gold Nanorods Enhance Light Emission. *ACS Nano* **2020**.
16. Cortés, E.; Xie, W.; Cambiasso, J.; Jermyn, A. S.; Sundararaman, R.; Narang, P.; Schlücker, S.; Maier, S. A., Plasmonic hot electron transport drives nano-localized chemistry. *Nature Communications* **2017,** *8* (1), 14880.
17. Yuan, L.; Lou, M.; Clark, B. D.; Lou, M.; Zhou, L.; Tian, S.; Jacobson, C. R.; Nordlander, P.; Halas, N. J., Morphology-Dependent Reactivity of a Plasmonic Photocatalyst. *ACS Nano* **2020,** *14* (9), 12054-12063.
18. Sousa-Castillo, A.; Comesaña-Hermo, M.; Rodríguez-González, B.; Pérez-Lorenzo, M.; Wang, Z.; Kong, X.-T.; Govorov, A. O.; Correa-Duarte, M. A., Boosting Hot Electron-Driven Photocatalysis through Anisotropic Plasmonic Nanoparticles with Hot Spots in Au–TiO2 Nanoarchitectures. *The Journal of Physical Chemistry C* **2016,** *120* (21), 11690-11699.
19. Chang, L.; Besteiro, L. V.; Sun, J.; Santiago, E. Y.; Gray, S. K.; Wang, Z.; Govorov, A. O., Electronic Structure of the Plasmons in Metal Nanocrystals: Fundamental Limitations for the Energy Efficiency of Hot Electron Generation. *ACS Energy Letters* **2019,** *4* (10), 2552-2568.

20. Ladstädter, F.; Hohenester, U.; Puschnig, P.; Ambrosch-Draxl, C., First-principles calculation of hot-electron scattering in metals. *Physical Review B* **2004,** *70* (23), 235125.
21. Besteiro, L. V.; Kong, X.-T.; Wang, Z.; Hartland, G.; Govorov, A. O., Understanding Hot-Electron Generation and Plasmon Relaxation in Metal Nanocrystals: Quantum and Classical Mechanisms. *ACS Photonics* **2017,** *4* (11), 2759-2781.
22. Leenheer, A. J.; Narang, P.; Lewis, N. S.; Atwater, H. A., Solar energy conversion via hot electron internal photoemission in metallic nanostructures: Efficiency estimates. *Journal of Applied Physics* **2014,** *115* (13), 134301.
23. Narang, P.; Sundararaman, R.; Atwater, H. A., Plasmonic hot carrier dynamics in solid-state and chemical systems for energy conversion *Nanophotonics* **2016,** *5* (1), 96.
24. Wen, L.; Xu, R.; Cui, C.; Tang, W.; Mi, Y.; Lu, X.; Zeng, Z.; Suib, S. L.; Gao, P.-X.; Lei, Y., Template-Guided Programmable Janus Heteronanostructure Arrays for Efficient Plasmonic Photocatalysis. *Nano Letters* **2018,** *18* (8), 4914-4921.
25. Tada, H.; Mitsui, T.; Kiyonaga, T.; Akita, T.; Tanaka, K., All-solid-state Z-scheme in CdS–Au–$TiO_2$ three-component nanojunction system. *Nature Materials* **2006,** *5* (10), 782-786.
26. Mubeen, S.; Lee, J.; Singh, N.; Krämer, S.; Stucky, G. D.; Moskovits, M., An autonomous photosynthetic device in which all charge carriers derive from surface plasmons. *Nature Nanotechnology* **2013,** *8* (4), 247-251.
27. Xiao, F.-X.; Zeng, Z.; Liu, B., Bridging the Gap: Electron Relay and Plasmonic Sensitization of Metal Nanocrystals for Metal Clusters. *Journal of the American Chemical Society* **2015,** *137* (33), 10735-10744.
28. Cai, X.; Chen, Q.; Wang, R.; Wang, A.; Wang, J.; Zhong, S.; Liu, Y.; Chen, J.; Bai, S., Integration of Plasmonic Metal and Cocatalyst: An Efficient Strategy for Boosting the Visible and Broad-Spectrum Photocatalytic $H_2$ Evolution. *Advanced Materials Interfaces* **2019,** *6* (17), 1900775.
29. Mascaretti, L.; Dutta, A.; Kment, Š.; Shalaev, V. M.; Boltasseva, A.; Zbořil, R.; Naldoni, A., Plasmon-Enhanced Photoelectrochemical Water Splitting for Efficient Renewable Energy Storage. *Advanced Materials* **2019,** *31* (31), 1805513.
30. Santiago, E. Y.; Besteiro, L. V.; Kong, X.-T.; Correa-Duarte, M. A.; Wang, Z.; Govorov, A. O., Efficiency of Hot-Electron Generation in Plasmonic Nanocrystals with Complex Shapes: Surface-Induced Scattering, Hot Spots, and Interband Transitions. *ACS Photonics* **2020**.
31. Lee, S.; Kim, J.; Yang, H.; Cortés, E.; Kang, S.; Han, S. W., Particle-in-a-Frame Nanostructures with Interior Nanogaps. *Angew. Chem. Int. Ed.* **2019,** *58* (44), 15890-15894.
32. Wu, K.; Chen, J.; McBride, J. R.; Lian, T., Efficient hot-electron transfer by a plasmon-induced interfacial charge-transfer transition. *Science* **2015,** *349* (6248), 632-635.
33. Kumar, P. V.; Rossi, T. P.; Marti-Dafcik, D.; Reichmuth, D.; Kuisma, M.; Erhart, P.; Puska, M. J.; Norris, D. J., Plasmon-Induced Direct Hot-Carrier Transfer at Metal–Acceptor Interfaces. *ACS Nano* **2019,** *13* (3), 3188-3195.
34. Sundararaman, R.; Narang, P.; Jermyn, A. S.; Goddard Iii, W. A.; Atwater, H. A., Theoretical predictions for hot-carrier generation from surface plasmon decay. *Nature Communications* **2014,** *5* (1), 5788.
35. Jermyn, A. S.; Tagliabue, G.; Atwater, H. A.; Goddard, W. A.; Narang, P.; Sundararaman, R., Transport of hot carriers in plasmonic nanostructures. *Physical Review Materials* **2019,** *3* (7), 075201.
36. Tagliabue, G.; DuChene, J. S.; Habib, A.; Sundararaman, R.; Atwater, H. A., Hot-Hole versus Hot-Electron Transport at Cu/GaN Heterojunction Interfaces. *ACS Nano* **2020,** *14* (5), 5788-5797.
37. Manjavacas, A.; Liu, J. G.; Kulkarni, V.; Nordlander, P., Plasmon-Induced Hot Carriers in Metallic Nanoparticles. *ACS Nano* **2014,** *8* (8), 7630-7638.
38. Douglas-Gallardo, O. A.; Berdakin, M.; Frauenheim, T.; Sánchez, C. G., Plasmon-induced hot-carrier generation differences in gold and silver nanoclusters. *Nanoscale* **2019,** *11* (17), 8604-8615.
39. Besteiro, L. V.; Yu, P.; Wang, Z.; Holleitner, A. W.; Hartland, G. V.; Wiederrecht, G. P.; Govorov, A. O., The fast and the furious: Ultrafast hot electrons in plasmonic metastructures. Size and structure matter. *Nano Today* **2019,** *27*, 120-145.

40. Hartland, G. V.; Besteiro, L. V.; Johns, P.; Govorov, A. O., What's so Hot about Electrons in Metal Nanoparticles? *ACS Energy Letters* **2017**, *2* (7), 1641-1653.
41. Brown, A. M.; Sundararaman, R.; Narang, P.; Schwartzberg, A. M.; Goddard, W. A.; Atwater, H. A., Experimental and Ab Initio Ultrafast Carrier Dynamics in Plasmonic Nanoparticles. *Physical Review Letters* **2017**, *118* (8), 087401.
42. Brown, A. M.; Sundararaman, R.; Narang, P.; Goddard, W. A.; Atwater, H. A., Ab initio phonon coupling and optical response of hot electrons in plasmonic metals. *Physical Review B* **2016**, *94* (7), 075120.
43. Su, M.-N.; Ciccarino, C. J.; Kumar, S.; Dongare, P. D.; Hosseini Jebeli, S. A.; Renard, D.; Zhang, Y.; Ostovar, B.; Chang, W.-S.; Nordlander, P.; Halas, N. J.; Sundararaman, R.; Narang, P.; Link, S., Ultrafast Electron Dynamics in Single Aluminum Nanostructures. *Nano Letters* **2019**, *19* (5), 3091-3097.
44. Ratchford, D. C.; Dunkelberger, A. D.; Vurgaftman, I.; Owrutsky, J. C.; Pehrsson, P. E., Quantification of Efficient Plasmonic Hot-Electron Injection in Gold Nanoparticle–TiO2 Films. *Nano Letters* **2017**, *17* (10), 6047-6055.
45. Codrington, J.; Eldabagh, N.; Fernando, K.; Foley, J. J., Unique Hot Carrier Distributions from Scattering-Mediated Absorption. *ACS Photonics* **2017**, *4* (3), 552-559.
46. Tagliabue, G.; DuChene, J. S.; Abdellah, M.; Habib, A.; Gosztola, D. J.; Hattori, Y.; Cheng, W.-H.; Zheng, K.; Canton, S. E.; Sundararaman, R.; Sá, J.; Atwater, H. A., Ultrafast hot-hole injection modifies hot-electron dynamics in Au/p-GaN heterostructures. *Nature Materials* **2020**.
47. Bernardi, M.; Mustafa, J.; Neaton, J. B.; Louie, S. G., Theory and computation of hot carriers generated by surface plasmon polaritons in noble metals. *Nature Communications* **2015**, *6* (1), 7044.
48. Lozan, O.; Sundararaman, R.; Ea-Kim, B.; Rampnoux, J.-M.; Narang, P.; Dilhaire, S.; Lalanne, P., Increased rise time of electron temperature during adiabatic plasmon focusing. *Nature Communications* **2017**, *8* (1), 1656.
49. Kim, Y.; Smith, J. G.; Jain, P. K., Harvesting multiple electron–hole pairs generated through plasmonic excitation of Au nanoparticles. *Nature Chemistry* **2018**, *10* (7), 763-769.
50. Khorashad, L. K.; Besteiro, L. V.; Correa-Duarte, M. A.; Burger, S.; Wang, Z. M.; Govorov, A. O., Hot Electrons Generated in Chiral Plasmonic Nanocrystals as a Mechanism for Surface Photochemistry and Chiral Growth. *Journal of the American Chemical Society* **2020**, *142* (9), 4193-4205.
51. Kong, X.-T.; Khosravi Khorashad, L.; Wang, Z.; Govorov, A. O., Photothermal Circular Dichroism Induced by Plasmon Resonances in Chiral Metamaterial Absorbers and Bolometers. *Nano Letters* **2018**, *18* (3), 2001-2008.
52. Saito, K.; Tatsuma, T., Chiral Plasmonic Nanostructures Fabricated by Circularly Polarized Light. *Nano Letters* **2018**, *18* (5), 3209-3212.
53. Kamarudheen, R.; Aalbers, G. J. W.; Hamans, R. F.; Kamp, L. P. J.; Baldi, A., Distinguishing Among All Possible Activation Mechanisms of a Plasmon-Driven Chemical Reaction. *ACS Energy Letters* **2020**, *5* (8), 2605-2613.
54. Sarhan, R. M.; Koopman, W.; Schuetz, R.; Schmid, T.; Liebig, F.; Koetz, J.; Bargheer, M., The importance of plasmonic heating for the plasmon-driven photodimerization of 4-nitrothiophenol. *Scientific Reports* **2019**, *9* (1), 3060.
55. Pensa, E.; Gargiulo, J.; Lauri, A.; Schlücker, S.; Cortés, E.; Maier, S. A., Spectral Screening of the Energy of Hot Holes over a Particle Plasmon Resonance. *Nano Letters* **2019**, *19* (3), 1867-1874.
56. Kamarudheen, R.; Kumari, G.; Baldi, A., Plasmon-driven synthesis of individual metal@semiconductor core@shell nanoparticles. *Nature Communications* **2020**, *11* (1), 3957.
57. Robert, H. M. L.; Kundrat, F.; Bermúdez-Ureña, E.; Rigneault, H.; Monneret, S.; Quidant, R.; Polleux, J.; Baffou, G., Light-Assisted Solvothermal Chemistry Using Plasmonic Nanoparticles. *ACS Omega* **2016**, *1* (1), 2-8.

58. Cao, L.; Barsic, D. N.; Guichard, A. R.; Brongersma, M. L., Plasmon-Assisted Local Temperature Control to Pattern Individual Semiconductor Nanowires and Carbon Nanotubes. *Nano Letters* **2007,** *7* (11), 3523-3527.
59. Barella, M.; Violi, I. L.; Gargiulo, J.; Martinez, L. P.; Goschin, F.; Guglielmotti, V.; Pallarola, D.; Schlücker, S.; Pilo-Pais, M.; Acuna, G. P.; Maier, S. A.; Cortés, E.; Stefani, F. D., In Situ Photothermal Response of Single Gold Nanoparticles through Hyperspectral Imaging Anti-Stokes Thermometry. *ACS Nano* **2020**.
60. Simoncelli, S.; Pensa, E. L.; Brick, T.; Gargiulo, J.; Lauri, A.; Cambiasso, J.; Li, Y.; Maier, S. A.; Cortés, E., Monitoring plasmonic hot-carrier chemical reactions at the single particle level. *Faraday Discussions* **2019,** *214* (0), 73-87.
61. Baffou, G.; Quidant, R.; García de Abajo, F. J., Nanoscale Control of Optical Heating in Complex Plasmonic Systems. *ACS Nano* **2010,** *4* (2), 709-716.
62. Baffou, G.; Berto, P.; Bermúdez Ureña, E.; Quidant, R.; Monneret, S.; Polleux, J.; Rigneault, H., Photoinduced Heating of Nanoparticle Arrays. *ACS Nano* **2013,** *7* (8), 6478-6488.
63. Hogan, N. J.; Urban, A. S.; Ayala-Orozco, C.; Pimpinelli, A.; Nordlander, P.; Halas, N. J., Nanoparticles Heat through Light Localization. *Nano Letters* **2014,** *14* (8), 4640-4645.
64. Kamarudheen, R.; Castellanos, G. W.; Kamp, L. P. J.; Clercx, H. J. H.; Baldi, A., Quantifying Photothermal and Hot Charge Carrier Effects in Plasmon-Driven Nanoparticle Syntheses. *ACS Nano* **2018,** *12* (8), 8447-8455.
65. Neumann, O.; Urban, A. S.; Day, J.; Lal, S.; Nordlander, P.; Halas, N. J., Solar Vapor Generation Enabled by Nanoparticles. *ACS Nano* **2013,** *7* (1), 42-49.
66. Dongare, P. D.; Alabastri, A.; Pedersen, S.; Zodrow, K. R.; Hogan, N. J.; Neumann, O.; Wu, J.; Wang, T.; Deshmukh, A.; Elimelech, M.; Li, Q.; Nordlander, P.; Halas, N. J., Nanophotonics-enabled solar membrane distillation for off-grid water purification. *PNAS* **2017,** *114* (27), 6936-6941.
67. Coventry, J.; Burge, P., Optical properties of Pyromark 2500 coatings of variable thicknesses on a range of materials for concentrating solar thermal applications. *AIP Conference Proceedings* **2017,** *1850* (1), 030012.
68. Naldoni, A.; Kudyshev, Z. A.; Mascaretti, L.; Sarmah, S. P.; Rej, S.; Froning, J. P.; Tomanec, O.; Yoo, J. E.; Wang, D.; Kment, Š.; Montini, T.; Fornasiero, P.; Shalaev, V. M.; Schmuki, P.; Boltasseva, A.; Zbořil, R., Solar Thermoplasmonic Nanofurnace for High-Temperature Heterogeneous Catalysis. *Nano Letters* **2020,** *20* (5), 3663-3672.
69. Alabastri, A., Flow-Driven Resonant Energy Systems. *Physical Review Applied* **2020,** *14* (3), 034045.
70. Alabastri, A.; Dongare, P. D.; Neumann, O.; Metz, J.; Adebiyi, I.; Nordlander, P.; Halas, N. J., Resonant energy transfer enhances solar thermal desalination. *Energy & Environmental Science* **2020,** *13* (3), 968-976.
71. Baffou, G.; Rigneault, H., Femtosecond-pulsed optical heating of gold nanoparticles. *Physical Review B* **2011,** *84* (3), 035415.
72. Alabastri, A.; Toma, A.; Malerba, M.; De Angelis, F.; Proietti Zaccaria, R., High Temperature Nanoplasmonics: The Key Role of Nonlinear Effects. *ACS Photonics* **2015,** *2* (1), 115-120.
73. Alabastri, A.; Malerba, M.; Calandrini, E.; Manjavacas, A.; De Angelis, F.; Toma, A.; Proietti Zaccaria, R., Controlling the Heat Dissipation in Temperature-Matched Plasmonic Nanostructures. *Nano Letters* **2017,** *17* (9), 5472-5480.
74. Christopher, P.; Xin, H.; Marimuthu, A.; Linic, S., Singular characteristics and unique chemical bond activation mechanisms of photocatalytic reactions on plasmonic nanostructures. *Nature Materials* **2012,** *11* (12), 1044-1050.
75. Zhang, X.; Li, X.; Zhang, D.; Su, N. Q.; Yang, W.; Everitt, H. O.; Liu, J., Product selectivity in plasmonic photocatalysis for carbon dioxide hydrogenation. *Nature Communications* **2017,** *8* (1), 14542.
76. Khurgin, J. B., Hot carriers generated by plasmons: where are they generated and where do they go from there? *Faraday Discussions* **2019,** *214* (0), 35-58.

77. Mukherjee, S.; Libisch, F.; Large, N.; Neumann, O.; Brown, L. V.; Cheng, J.; Lassiter, J. B.; Carter, E. A.; Nordlander, P.; Halas, N. J., Hot Electrons Do the Impossible: Plasmon-Induced Dissociation of H2 on Au. *Nano Letters* **2013,** *13* (1), 240-247.
78. Mukherjee, S.; Zhou, L.; Goodman, A. M.; Large, N.; Ayala-Orozco, C.; Zhang, Y.; Nordlander, P.; Halas, N. J., Hot-Electron-Induced Dissociation of H2 on Gold Nanoparticles Supported on SiO2. *Journal of the American Chemical Society* **2014,** *136* (1), 64-67.
79. Zhou, L.; Swearer, D. F.; Zhang, C.; Robatjazi, H.; Zhao, H.; Henderson, L.; Dong, L.; Christopher, P.; Carter, E. A.; Nordlander, P.; Halas, N. J., Quantifying hot carrier and thermal contributions in plasmonic photocatalysis. *Science* **2018,** *362* (6410), 69-72.
80. Hu, C.; Chen, X.; Jin, J.; Han, Y.; Chen, S.; Ju, H.; Cai, J.; Qiu, Y.; Gao, C.; Wang, C.; Qi, Z.; Long, R.; Song, L.; Liu, Z.; Xiong, Y., Surface Plasmon Enabling Nitrogen Fixation in Pure Water through a Dissociative Mechanism under Mild Conditions. *Journal of the American Chemical Society* **2019,** *141* (19), 7807-7814.
81. Dubi, Y.; Un, I. W.; Sivan, Y., Thermal effects – an alternative mechanism for plasmon-assisted photocatalysis. *Chemical Science* **2020,** *11* (19), 5017-5027.
82. Jain, P. K., Comment on "Thermal effects – an alternative mechanism for plasmon-assisted photocatalysis" by Y. Dubi, I. W. Un and Y. Sivan, Chem. Sci., 2020, 11, 5017. *Chemical Science* **2020,** *11* (33), 9022-9023.
83. Baffou, G.; Bordacchini, I.; Baldi, A.; Quidant, R., Simple experimental procedures to distinguish photothermal from hot-carrier processes in plasmonics. *Light: Science & Applications* **2020,** *9* (1), 108.
84. Richardson, H. H.; Carlson, M. T.; Tandler, P. J.; Hernandez, P.; Govorov, A. O., Experimental and Theoretical Studies of Light-to-Heat Conversion and Collective Heating Effects in Metal Nanoparticle Solutions. *Nano Letters* **2009,** *9* (3), 1139-1146.
85. Zhang, X.; Li, X.; Reish, M. E.; Zhang, D.; Su, N. Q.; Gutiérrez, Y.; Moreno, F.; Yang, W.; Everitt, H. O.; Liu, J., Plasmon-Enhanced Catalysis: Distinguishing Thermal and Nonthermal Effects. *Nano Letters* **2018,** *18* (3), 1714-1723.
86. Sivan, Y.; Baraban, J.; Un, I. W.; Dubi, Y., Comment on "Quantifying hot carrier and thermal contributions in plasmonic photocatalysis". *Science* **2019,** *364* (6439), eaaw9367.
87. Zhou, L.; Swearer, D. F.; Robatjazi, H.; Alabastri, A.; Christopher, P.; Carter, E. A.; Nordlander, P.; Halas, N. J., Response to Comment on "Quantifying hot carrier and thermal contributions in plasmonic photocatalysis". *Science* **2019,** *364* (6439), eaaw9545.
88. Baffou, G.; Quidant, R., Nanoplasmonics for chemistry. *Chemical Society Reviews* **2014,** *43* (11), 3898-3907.
89. Boerigter, C.; Campana, R.; Morabito, M.; Linic, S., Evidence and implications of direct charge excitation as the dominant mechanism in plasmon-mediated photocatalysis. *Nature Communications* **2016,** *7* (1), 10545.
90. Rodio, M.; Graf, M.; Schulz, F.; Mueller, N. S.; Eich, M.; Lange, H., Experimental Evidence for Nonthermal Contributions to Plasmon-Enhanced Electrochemical Oxidation Reactions. *ACS Catalysis* **2020,** *10* (3), 2345-2353.
91. Ou, W.; Zhou, B.; Shen, J.; Lo, T. W.; Lei, D.; Li, S.; Zhong, J.; Li, Y. Y.; Lu, J., Thermal and Nonthermal Effects in Plasmon-Mediated Electrochemistry at Nanostructured Ag Electrodes. **2020,** *59* (17), 6790-6793.
92. Christopher, P.; Xin, H.; Linic, S., Visible-light-enhanced catalytic oxidation reactions on plasmonic silver nanostructures. *Nature Chemistry* **2011,** *3* (6), 467-472.
93. Yu, S.; Jain, P. K., Isotope Effects in Plasmonic Photosynthesis. *Angew. Chem. Int. Ed.* **2020,** *n/a* (n/a).
94. Keller, E. L.; Frontiera, R. R., Ultrafast Nanoscale Raman Thermometry Proves Heating Is Not a Primary Mechanism for Plasmon-Driven Photocatalysis. *ACS Nano* **2018,** *12* (6), 5848-5855.

95. Yu, Y.; Sundaresan, V.; Willets, K. A., Hot Carriers versus Thermal Effects: Resolving the Enhancement Mechanisms for Plasmon-Mediated Photoelectrochemical Reactions. *The Journal of Physical Chemistry C* **2018,** *122* (9), 5040-5048.
96. Simoncelli, S.; Li, Y.; Cortés, E.; Maier, S. A., Nanoscale Control of Molecular Self-Assembly Induced by Plasmonic Hot-Electron Dynamics. *ACS Nano* **2018,** *12* (3), 2184-2192.
97. Quiroz, J.; Barbosa, E. C. M.; Araujo, T. P.; Fiorio, J. L.; Wang, Y.-C.; Zou, Y.-C.; Mou, T.; Alves, T. V.; de Oliveira, D. C.; Wang, B.; Haigh, S. J.; Rossi, L. M.; Camargo, P. H. C., Controlling Reaction Selectivity over Hybrid Plasmonic Nanocatalysts. *Nano Letters* **2018,** *18* (11), 7289-7297.
98. DuChene, J. S.; Tagliabue, G.; Welch, A. J.; Cheng, W.-H.; Atwater, H. A., Hot Hole Collection and Photoelectrochemical CO2 Reduction with Plasmonic Au/p-GaN Photocathodes. *Nano Letters* **2018,** *18* (4), 2545-2550.
99. DuChene, J. S.; Tagliabue, G.; Welch, A. J.; Li, X.; Cheng, W.-H.; Atwater, H. A., Optical Excitation of a Nanoparticle Cu/p-NiO Photocathode Improves Reaction Selectivity for CO2 Reduction in Aqueous Electrolytes. *Nano Letters* **2020,** *20* (4), 2348-2358.
100. Aizpurua, J.; Ashfold, M.; Baletto, F.; Baumberg, J.; Christopher, P.; Cortés, E.; de Nijs, B.; Diaz Fernandez, Y.; Gargiulo, J.; Gawinkowski, S.; Halas, N.; Hamans, R.; Jankiewicz, B.; Khurgin, J.; Kumar, P. V.; Liu, J.; Maier, S.; Maurer, R. J.; Mount, A.; Mueller, N. S.; Oulton, R.; Parente, M.; Park, J. Y.; Polanyi, J.; Quiroz, J.; Rejman, S.; Schlücker, S.; Schultz, Z.; Sivan, Y.; Tagliabue, G.; Thangamuthu, M.; Torrente-Murciano, L.; Xiao, X.; Zayats, A.; Zhan, C., Dynamics of hot electron generation in metallic nanostructures: general discussion. *Faraday Discussions* **2019,** *214* (0), 123-146.
101. Li, W.; Coppens, Z. J.; Besteiro, L. V.; Wang, W.; Govorov, A. O.; Valentine, J., Circularly polarized light detection with hot electrons in chiral plasmonic metamaterials. *Nature Communications* **2015,** *6* (1), 8379.
102. Tagliabue, G.; Jermyn, A. S.; Sundararaman, R.; Welch, A. J.; DuChene, J. S.; Pala, R.; Davoyan, A. R.; Narang, P.; Atwater, H. A., Quantifying the role of surface plasmon excitation and hot carrier transport in plasmonic devices. *Nature Communications* **2018,** *9* (1), 3394.
103. Furube, A.; Du, L.; Hara, K.; Katoh, R.; Tachiya, M., Ultrafast Plasmon-Induced Electron Transfer from Gold Nanodots into TiO2 Nanoparticles. *Journal of the American Chemical Society* **2007,** *129* (48), 14852-14853.
104. Zheng, B. Y.; Zhao, H.; Manjavacas, A.; McClain, M.; Nordlander, P.; Halas, N. J., Distinguishing between plasmon-induced and photoexcited carriers in a device geometry. *Nature Communications* **2015,** *6* (1), 7797.
105. Ng, C.; Zeng, P.; Lloyd, J. A.; Chakraborty, D.; Roberts, A.; Smith, T. A.; Bach, U.; Sader, J. E.; Davis, T. J.; Gómez, D. E., Large-Area Nanofabrication of Partially Embedded Nanostructures for Enhanced Plasmonic Hot-Carrier Extraction. *ACS Applied Nano Materials* **2019,** *2* (3), 1164-1169.
106. Grajower, M.; Levy, U.; Khurgin, J. B., The Role of Surface Roughness in Plasmonic-Assisted Internal Photoemission Schottky Photodetectors. *ACS Photonics* **2018,** *5* (10), 4030-4036.
107. Lee, H.; Lee, H.; Park, J. Y., Direct Imaging of Surface Plasmon-Driven Hot Electron Flux on the Au Nanoprism/TiO2. *Nano Letters* **2019,** *19* (2), 891-896.
108. Du, L.; Furube, A.; Hara, K.; Katoh, R.; Tachiya, M., Ultrafast plasmon induced electron injection mechanism in gold–TiO2 nanoparticle system. *Journal of Photochemistry and Photobiology C: Photochemistry Reviews* **2013,** *15*, 21-30.
109. Heilpern, T.; Manjare, M.; Govorov, A. O.; Wiederrecht, G. P.; Gray, S. K.; Harutyunyan, H., Determination of hot carrier energy distributions from inversion of ultrafast pump-probe reflectivity measurements. *Nature Communications* **2018,** *9* (1), 1853.
110. Yu, Y.; Wijesekara, K. D.; Xi, X.; Willets, K. A., Quantifying Wavelength-Dependent Plasmonic Hot Carrier Energy Distributions at Metal/Semiconductor Interfaces. *ACS Nano* **2019,** *13* (3), 3629-3637.
111. Reddy, H.; Wang, K.; Kudyshev, Z.; Zhu, L.; Yan, S.; Vezzoli, A.; Higgins, S. J.; Gavini, V.; Boltasseva, A.; Reddy, P.; Shalaev, V. M.; Meyhofer, E., Determining plasmonic hot-carrier energy distributions via single-molecule transport measurements. *Science* **2020,** *369* (6502), 423-426.

112. Sistani, M.; Bartmann, M. G.; Güsken, N. A.; Oulton, R. F.; Keshmiri, H.; Luong, M. A.; Momtaz, Z. S.; Den Hertog, M. I.; Lugstein, A., Plasmon-Driven Hot Electron Transfer at Atomically Sharp Metal–Semiconductor Nanojunctions. *ACS Photonics* **2020,** *7* (7), 1642-1648.
113. Tan, S.; Argondizzo, A.; Ren, J.; Liu, L.; Zhao, J.; Petek, H., Plasmonic coupling at a metal/semiconductor interface. *Nature Photonics* **2017,** *11* (12), 806-812.
114. Cushing, S. K.; Chen, C.-J.; Dong, C. L.; Kong, X.-T.; Govorov, A. O.; Liu, R.-S.; Wu, N., Tunable Nonthermal Distribution of Hot Electrons in a Semiconductor Injected from a Plasmonic Gold Nanostructure. *ACS Nano* **2018,** *12* (7), 7117-7126.
115. Foerster, B.; Hartelt, M.; Collins, S. S. E.; Aeschlimann, M.; Link, S.; Sönnichsen, C., Interfacial States Cause Equal Decay of Plasmons and Hot Electrons at Gold–Metal Oxide Interfaces. *Nano Letters* **2020,** *20* (5), 3338-3343.
116. Khurgin, J. B.; Levy, U., Generating Hot Carriers in Plasmonic Nanoparticles: When Quantization Does Matter? *ACS Photonics* **2020,** *7* (3), 547-553.
117. Narang, P.; Sundararaman, R.; Jermyn, A. S.; Goddard, W. A.; Atwater, H. A., Cubic Nonlinearity Driven Up-Conversion in High-Field Plasmonic Hot Carrier Systems. *The Journal of Physical Chemistry C* **2016,** *120* (37), 21056-21062.
118. Tomko, J. A.; Runnerstrom, E. L.; Wang, Y.-S.; Chu, W.; Nolen, J. R.; Olson, D. H.; Kelley, K. P.; Cleri, A.; Nordlander, J.; Caldwell, J. D.; Prezhdo, O. V.; Maria, J.-P.; Hopkins, P. E., Long-lived modulation of plasmonic absorption by ballistic thermal injection. *Nature Nanotechnology* **2020**.
119. Zhan, C.; Wang, Z.-Y.; Zhang, X.-G.; Chen, X.-J.; Huang, Y.-F.; Hu, S.; Li, J.-F.; Wu, D.-Y.; Moskovits, M.; Tian, Z.-Q., Interfacial Construction of Plasmonic Nanostructures for the Utilization of the Plasmon-Excited Electrons and Holes. *Journal of the American Chemical Society* **2019,** *141* (20), 8053-8057.
120. Cortés, E., Activating plasmonic chemistry. *Science* **2018,** *362* (6410), 28-29.
121. Aslam, U.; Chavez, S.; Linic, S., Controlling energy flow in multimetallic nanostructures for plasmonic catalysis. *Nature Nanotechnology* **2017,** *12* (10), 1000-1005.
122. Swearer, D. F.; Zhao, H.; Zhou, L.; Zhang, C.; Robatjazi, H.; Martirez, J. M. P.; Krauter, C. M.; Yazdi, S.; McClain, M. J.; Ringe, E.; Carter, E. A.; Nordlander, P.; Halas, N. J., Heterometallic antenna−reactor complexes for photocatalysis. *PNAS* **2016,** *113* (32), 8916-8920.
123. Jia, H.; Du, A.; Zhang, H.; Yang, J.; Jiang, R.; Wang, J.; Zhang, C.-y., Site-Selective Growth of Crystalline Ceria with Oxygen Vacancies on Gold Nanocrystals for Near-Infrared Nitrogen Photofixation. *Journal of the American Chemical Society* **2019,** *141* (13), 5083-5086.
124. Kazuma, E.; Lee, M.; Jung, J.; Trenary, M.; Kim, Y., Single-Molecule Study of a Plasmon-Induced Reaction for a Strongly Chemisorbed Molecule. *Angew. Chem. Int. Ed.* **2020,** *59* (20), 7960-7966.
125. Guo, J.; Zhang, Y.; Shi, L.; Zhu, Y.; Mideksa, M. F.; Hou, K.; Zhao, W.; Wang, D.; Zhao, M.; Zhang, X.; Lv, J.; Zhang, J.; Wang, X.; Tang, Z., Boosting Hot Electrons in Hetero-superstructures for Plasmon-Enhanced Catalysis. *Journal of the American Chemical Society* **2017,** *139* (49), 17964-17972.
126. Peljo, P.; Manzanares, J. A.; Girault, H. H., Contact Potentials, Fermi Level Equilibration, and Surface Charging. *Langmuir* **2016,** *32* (23), 5765-5775.
127. Engelbrekt, C.; Crampton, K. T.; Fishman, D. A.; Law, M.; Apkarian, V. A., Efficient Plasmon-Mediated Energy Funneling to the Surface of Au@Pt Core–Shell Nanocrystals. *ACS Nano* **2020,** *14* (4), 5061-5074.
128. Foerster, B.; Spata, V. A.; Carter, E. A.; Sönnichsen, C.; Link, S., Plasmon damping depends on the chemical nature of the nanoparticle interface. *Science Advances* **2019,** *5* (3), eaav0704.
129. Li, K.; Hogan, N. J.; Kale, M. J.; Halas, N. J.; Nordlander, P.; Christopher, P., Balancing Near-Field Enhancement, Absorption, and Scattering for Effective Antenna–Reactor Plasmonic Photocatalysis. *Nano Letters* **2017,** *17* (6), 3710-3717.
130. Rao, V. G.; Aslam, U.; Linic, S., Chemical Requirement for Extracting Energetic Charge Carriers from Plasmonic Metal Nanoparticles to Perform Electron-Transfer Reactions. *Journal of the American Chemical Society* **2019,** *141* (1), 643-647.

131. Xie, W.; Schlücker, S., Hot electron-induced reduction of small molecules on photorecycling metal surfaces. *Nature Communications* **2015,** *6* (1), 7570.
132. Chavez, S.; Aslam, U.; Linic, S., Design Principles for Directing Energy and Energetic Charge Flow in Multicomponent Plasmonic Nanostructures. *ACS Energy Letters* **2018,** *3* (7), 1590-1596.
133. Yu, S.; Wilson, A. J.; Heo, J.; Jain, P. K., Plasmonic Control of Multi-Electron Transfer and C–C Coupling in Visible-Light-Driven $CO_2$ Reduction on Au Nanoparticles. *Nano Letters* **2018,** *18* (4), 2189-2194.
134. Qi, J.; Resasco, J.; Robatjazi, H.; Alvarez, I. B.; Abdelrahman, O.; Dauenhauer, P.; Christopher, P., Dynamic Control of Elementary Step Energetics via Pulsed Illumination Enhances Photocatalysis on Metal Nanoparticles. *ACS Energy Letters* **2020**, 3518-3525.
135. Cortés, E., Efficiency and Bond Selectivity in Plasmon-Induced Photochemistry. *Advanced Optical Materials* **2017,** *5* (15), 1700191.
136. Hens, Z.; De Roo, J., Atomically Precise Nanocrystals. *Journal of the American Chemical Society* **2020,** *142* (37), 15627-15637.
137. Han, L.; Zhang, L.; Wu, H.; Zu, H.; Cui, P.; Guo, J.; Guo, R.; Ye, J.; Zhu, J.; Zheng, X.; Yang, L.; Zhong, Y.; Liang, S.; Wang, L., Anchoring Pt Single Atoms on Te Nanowires for Plasmon-Enhanced Dehydrogenation of Formic Acid at Room Temperature. *Advanced Science* **2019,** *6* (12), 1900006.
138. Zhou, L.; Martirez, J. M. P.; Finzel, J.; Zhang, C.; Swearer, D. F.; Tian, S.; Robatjazi, H.; Lou, M.; Dong, L.; Henderson, L.; Christopher, P.; Carter, E. A.; Nordlander, P.; Halas, N. J., Light-driven methane dry reforming with single atomic site antenna-reactor plasmonic photocatalysts. *Nature Energy* **2020,** *5* (1), 61-70.
139. Benz, F.; Schmidt, M. K.; Dreismann, A.; Chikkaraddy, R.; Zhang, Y.; Demetriadou, A.; Carnegie, C.; Ohadi, H.; de Nijs, B.; Esteban, R.; Aizpurua, J.; Baumberg, J. J., Single-molecule optomechanics in "picocavities". **2016,** *354* (6313), 726-729.
140. Mertens, J.; Demetriadou, A.; Bowman, R. W.; Benz, F.; Kleemann, M. E.; Tserkezis, C.; Shi, Y.; Yang, H. Y.; Hess, O.; Aizpurua, J.; Baumberg, J. J., Tracking Optical Welding through Groove Modes in Plasmonic Nanocavities. *Nano Letters* **2016,** *16* (9), 5605-5611.
141. Baumberg, J. J.; Aizpurua, J.; Mikkelsen, M. H.; Smith, D. R., Extreme nanophotonics from ultrathin metallic gaps. *Nature Materials* **2019,** *18* (7), 668-678.
142. Lee, J.; Crampton, K. T.; Tallarida, N.; Apkarian, V. A., Visualizing vibrational normal modes of a single molecule with atomically confined light. *Nature* **2019,** *568* (7750), 78-82.
143. Zhang, Y.; Yang, B.; Ghafoor, A.; Zhang, Y.; Zhang, Y.-F.; Wang, R.-P.; Yang, J.-L.; Luo, Y.; Dong, Z.-C.; Hou, J. G., Visually constructing the chemical structure of a single molecule by scanning Raman picoscopy. *National Science Review* **2019,** *6* (6), 1169-1175.
144. Yang, B.; Chen, G.; Ghafoor, A.; Zhang, Y.; Zhang, Y.; Zhang, Y.; Luo, Y.; Yang, J.; Sandoghdar, V.; Aizpurua, J.; Dong, Z.; Hou, J. G., Sub-nanometre resolution in single-molecule photoluminescence imaging. *Nature Photonics* **2020**.
145. Zhang, Y.; Dong, Z.-C.; Aizpurua, J., Theoretical treatment of single-molecule scanning Raman picoscopy in strongly inhomogeneous near fields. *Journal of Raman Spectroscopy* **2020,** *n/a* (n/a).
146. Li, C.-Y.; Duan, S.; Yi, J.; Wang, C.; Radjenovic, P. M.; Tian, Z.-Q.; Li, J.-F., Real-time detection of single-molecule reaction by plasmon-enhanced spectroscopy. *Science Advances* **2020,** *6* (24), eaba6012.
147. Sprague-Klein, E. A.; McAnally, M. O.; Zhdanov, D. V.; Zrimsek, A. B.; Apkarian, V. A.; Seideman, T.; Schatz, G. C.; Van Duyne, R. P., Observation of Single Molecule Plasmon-Driven Electron Transfer in Isotopically Edited 4,4′-Bipyridine Gold Nanosphere Oligomers. *Journal of the American Chemical Society* **2017,** *139* (42), 15212-15221.
148. Choi, H.-K.; Park, W.-H.; Park, C.-G.; Shin, H.-H.; Lee, K. S.; Kim, Z. H., Metal-Catalyzed Chemical Reaction of Single Molecules Directly Probed by Vibrational Spectroscopy. *Journal of the American Chemical Society* **2016,** *138* (13), 4673-4684.

149. de Nijs, B.; Benz, F.; Barrow, S. J.; Sigle, D. O.; Chikkaraddy, R.; Palma, A.; Carnegie, C.; Kamp, M.; Sundararaman, R.; Narang, P.; Scherman, O. A.; Baumberg, J. J., Plasmonic tunnel junctions for single-molecule redox chemistry. *Nature Communications* **2017,** *8* (1), 994.

150. Zhang, R.; Zhang, Y.; Dong, Z. C.; Jiang, S.; Zhang, C.; Chen, L. G.; Zhang, L.; Liao, Y.; Aizpurua, J.; Luo, Y.; Yang, J. L.; Hou, J. G., Chemical mapping of a single molecule by plasmon-enhanced Raman scattering. *Nature* **2013,** *498* (7452), 82-86.

151. Cortés, E.; Etchegoin, P. G.; Le Ru, E. C.; Fainstein, A.; Vela, M. E.; Salvarezza, R. C., Monitoring the Electrochemistry of Single Molecules by Surface-Enhanced Raman Spectroscopy. *Journal of the American Chemical Society* **2010,** *132* (51), 18034-18037.

152. Cortés, E.; Etchegoin, P. G.; Le Ru, E. C.; Fainstein, A.; Vela, M. E.; Salvarezza, R. C., Strong Correlation between Molecular Configurations and Charge-Transfer Processes Probed at the Single-Molecule Level by Surface-Enhanced Raman Scattering. *Journal of the American Chemical Society* **2013,** *135* (7), 2809-2815.

153. Chikkaraddy, R.; de Nijs, B.; Benz, F.; Barrow, S. J.; Scherman, O. A.; Rosta, E.; Demetriadou, A.; Fox, P.; Hess, O.; Baumberg, J. J., Single-molecule strong coupling at room temperature in plasmonic nanocavities. *Nature* **2016,** *535* (7610), 127-130.

154. Flick, J.; Rivera, N.; Narang, P., Strong light-matter coupling in quantum chemistry and quantum photonics *Nanophotonics* **2018,** *7* (9), 1479.

155. Thomas, A.; Lethuillier-Karl, L.; Nagarajan, K.; Vergauwe, R. M. A.; George, J.; Chervy, T.; Shalabney, A.; Devaux, E.; Genet, C.; Moran, J.; Ebbesen, T. W., Tilting a ground-state reactivity landscape by vibrational strong coupling. *Science* **2019,** *363* (6427), 615-619.

156. Pang, Y.; Thomas, A.; Nagarajan, K.; Vergauwe, R. M. A.; Joseph, K.; Patrahau, B.; Wang, K.; Genet, C.; Ebbesen, T. W., On the Role of Symmetry in Vibrational Strong Coupling: The Case of Charge-Transfer Complexation. *Angew. Chem. Int. Ed.* **2020,** *59* (26), 10436-10440.

157. Flick, J.; Ruggenthaler, M.; Appel, H.; Rubio, A., Atoms and molecules in cavities, from weak to strong coupling in quantum-electrodynamics (QED) chemistry. *PNAS* **2017,** *114* (12), 3026-3034.

158. Tokatly, I. V., Time-Dependent Density Functional Theory for Many-Electron Systems Interacting with Cavity Photons. *Physical Review Letters* **2013,** *110* (23), 233001.

159. Flick, J.; Narang, P., Cavity-Correlated Electron-Nuclear Dynamics from First Principles. *Physical Review Letters* **2018,** *121* (11), 113002.

160. Rivera, N.; Flick, J.; Narang, P., Variational Theory of Nonrelativistic Quantum Electrodynamics. *Physical Review Letters* **2019,** *122* (19), 193603.

*Suggested pull quotes*

By presenting comprehensive temperature- and frequency-dependent predictions of the optical response of hot electrons in plasmonic metals, theoretical and computational results have opened new avenues in plasmonic catalysis.

Even small uncertainties in the determination of the temperature of the catalytic active sites can result in large over- or under-estimations of the reaction rate enhancement due to nonthermal plasmonic effects.

Processes driven by hot charge carriers are characteristically "resonant" because they strongly depend on the excitation wavelength. On the contrary, photothermal effects are dissipative and only depend on the total absorbed optical power in the NP catalysts.

Thorough modeling and experimental research are critical to clarify the effects of charge separation across a metal/semiconductor interface on the hot-carrier population that remains available for photocatalysis.

The possibility of using sunlight to alter the activation barrier of certain chemical reaction pathways is undoubtedly a major step toward controlling chemical reactivity by external means.

Although thus far picocavities have been used to access vibrational modes at the submolecular level, they have the potential to offer extraordinary control over catalytic reactions at the single-molecule level as well.

From a theoretical perspective, it is particularly exciting to note the intersection between plasmonic nanochemistry and the work in strong-coupling chemistry.